\documentclass[
 aip,
rsi,
 amsmath,amssymb,
 reprint,
]{revtex4-1}

\usepackage{graphicx}\usepackage{dcolumn}
\usepackage{bm}
\usepackage[export]{adjustbox}
\usepackage{units}
\usepackage{gensymb}
\usepackage{siunitx}
\begin{document}

\preprint{AIP/1
3-QED}

\title{Determination of the resistivity anisotropy of orthorhombic materials via transverse resistivity measurements}

\author{P. Walmsley}
 \email{pwalms@stanford.edu}
 \affiliation{ 
Geballe Laboratory for Advanced Materials and Department of Applied Physics,\\ Stanford University, Stanford, California 94305-4045, USA
}
 \affiliation{Stanford Institute for Materials and Energy Sciences, SLAC National Accelerator Laboratory,\\ 2575 Sand Hill Road, Menlo Park, California 94025, USA
 }

\author{I. R. Fisher}
 \affiliation{ 
Geballe Laboratory for Advanced Materials and Department of Applied Physics,\\ Stanford University, Stanford, California 94305-4045, USA
}
 \affiliation{Stanford Institute for Materials and Energy Sciences, SLAC National Accelerator Laboratory,\\ 2575 Sand Hill Road, Menlo Park, California 94025, USA
 }

\date{\today}

\begin{abstract}
Measurements of the resistivity anisotropy can provide crucial information about the electronic structure and scattering processes in anisotropic and low-dimensional materials, but quantitative measurements by conventional means often suffer very significant systematic errors. Here we describe a novel approach to measuring the resistivity anisotropy of orthorhombic materials, using a single crystal and a single measurement, that is derived from a $\frac{\pi}{4}$ rotation of the measurement frame relative to the crystallographic axes. In this new basis the transverse resistivity gives a direct measurement of the resistivity anisotropy, which combined with the longitudinal resistivity also gives the in-plane elements of the conventional resistivity tensor via a 5-point contact geometry. This is demonstrated through application to the charge-density wave compound ErTe$_3$, and it is concluded that this method presents a significant improvement on existing techniques in many scenarios.
\end{abstract}

\pacs{Valid PACS appear here}
\maketitle

\section{\label{sec:introduction}Introduction}

Electrical transport measurements have long been a cornerstone of condensed matter physics as they are necessarily sensitive to the Fermi surface and interactions that are close in energy to the Fermi level. By extension, anisotropies in electrical transport can reflect the presence of broken symmetries, and hence their measurement can contribute to understanding the origin of novel phase transitions, particularly those that are driven by interactions at the Fermi-level. For example, in the case of an electronic nematic phase transition\cite{fradkin2009nematic,kivelson1998electronic}, close to the critical temperature the resistivity anisotropy is proportional to the nematic order parameter\cite{proportionalitycomment, carlson2006hysteresis, fernandes2011anisotropicres, blomberg2012effect}, motivating measurement of the resistivity anisotropy for detwinned samples. The Fe-based superconductors provide a recent example of such an effect. For several families of Fe-based materials, measurement of the resistivity anisotropy in the broken symmetry state\cite{fisher2011plane,chu2010anisotropy,man2015dai,tanatar2015fese,tanatar2010uniaxial,blomberg2011srfeas,kuo2011possible,liang2011effects,ying2011measurements,Jiang2012anisotropy}, and also measurement of the strain-induced resistivity anisotropy (elastoresistivity) in the tetragonal state\cite{chu2012nematic,kuo2016ubiquitous,watson2015fese,kuo2013measurement,kuo2014effect,shapiro2016measurement}, have provided compelling evidence that the tetragonal-to-orthorhombic phase transition that occurs in many of these materials is indeed driven by electronic correlations. Distinct from the previous example are materials that are fundamentally orthorhombic even at high temperatures, but which nevertheless develop an enhanced electronic anisotropy below some characteristic temperature. In these cases, changes in the resistivity anisotropy can still reveal important information regarding the origin of the associated phase transition or cross-over, also motivating measurement of the temperature dependence of the resistivity anisotropy. A well-known example is that of YBa$_2$Cu$_3$O$_{7-\delta}$, for which the presence of CuO chains leads to a fundamentally orthorhombic crystal structure, yet for which several measurements indicate the onset of an enhancement in the electronic anisotropy below a characteristic temperature, the physical origin of which remains poorly understood\cite{ando2002YBCO, cyr2015ybco,daou2010broken,chang2011nernst}. A second example in this latter class for which the physical origin is much clearer is the quasi 2-D material $R$Te$_3$ (where $R$ is a rare-earth ion). This material is also orthorhombic at high temperature (due to the presence of a glide plane in the crystallographic $a$ axis), but develops an increased anisotropy below the onset temperature of a uni-directional charge density wave (CDW) state\cite{sinchenko2014montgomery}. The specific case of ErTe$_3$ is further discussed below in the context of the present work.

Although the motivation to determine the resistivity anisotropy of orthorhombic materials like those mentioned above is often clear, the quantitative scope of resistivity anisotropy measurements can be significantly restricted by experimental limitations, particularly for small samples. In this paper, following a brief introduction to, and appraisal of, conventional techniques, we present a novel approach to directly measure the resistivity anisotropy via the transverse resistivity in a rotated experimental frame that addresses some of these limitations without invoking additional assumptions or instrumentation. The resistivity anisotropy of the $a-c$ plane in ErTe$_3$ is then presented to demonstrate the efficacy of the technique.
\vfill
\pagebreak
\section{\label{sec:methods}Methods to measure the resistivity anisotropy for an orthorhombic material}

\subsection{\label{sec:res_anis_def}Definition of resistivity anisotropy}

In an anisotropic material, the electrical resistivity is described by a second rank tensor $\boldsymbol{\rho}$ that relates the current density $J_j$ to the electric field $E_i$ via the relationship $E_i = \sum_i \rho_{ij} J_j$. When the orientation of the Cartesian basis is defined as parallel to the orthonormal crystallographic axes of the sample ($x\parallel a$, $y\parallel b$, $z\parallel c$), with some considerations of symmetry, this produces the conventional zero-field resistivity tensor,

\begin{equation}
\boldsymbol{\rho}=
\begin{pmatrix}
\rho_{a}&0&0\\
0&\rho_{b}&0\\
0&0&\rho_{c}\\
\end{pmatrix}
.
\label{eqn:resistivity_tensor}
\end{equation}

The resistivity anisotropy is then generally defined as the difference between two given diagonal components $(\rho_{ii}-\rho_{jj})$, although the figures of merit for the dimensionless resistivity anisotropy $\frac{\rho_{ii}-\rho_{jj}}{\nicefrac{1}{2}(\rho_{ii}+\rho_{jj})}$ and $\frac{\rho_{ii}}{\rho_{jj}}$ are often more meaningful quantities.

\subsection{Measurement of resistivity anisotropy by conventional methods}

\begin{figure} 
\includegraphics[width=0.88\columnwidth]{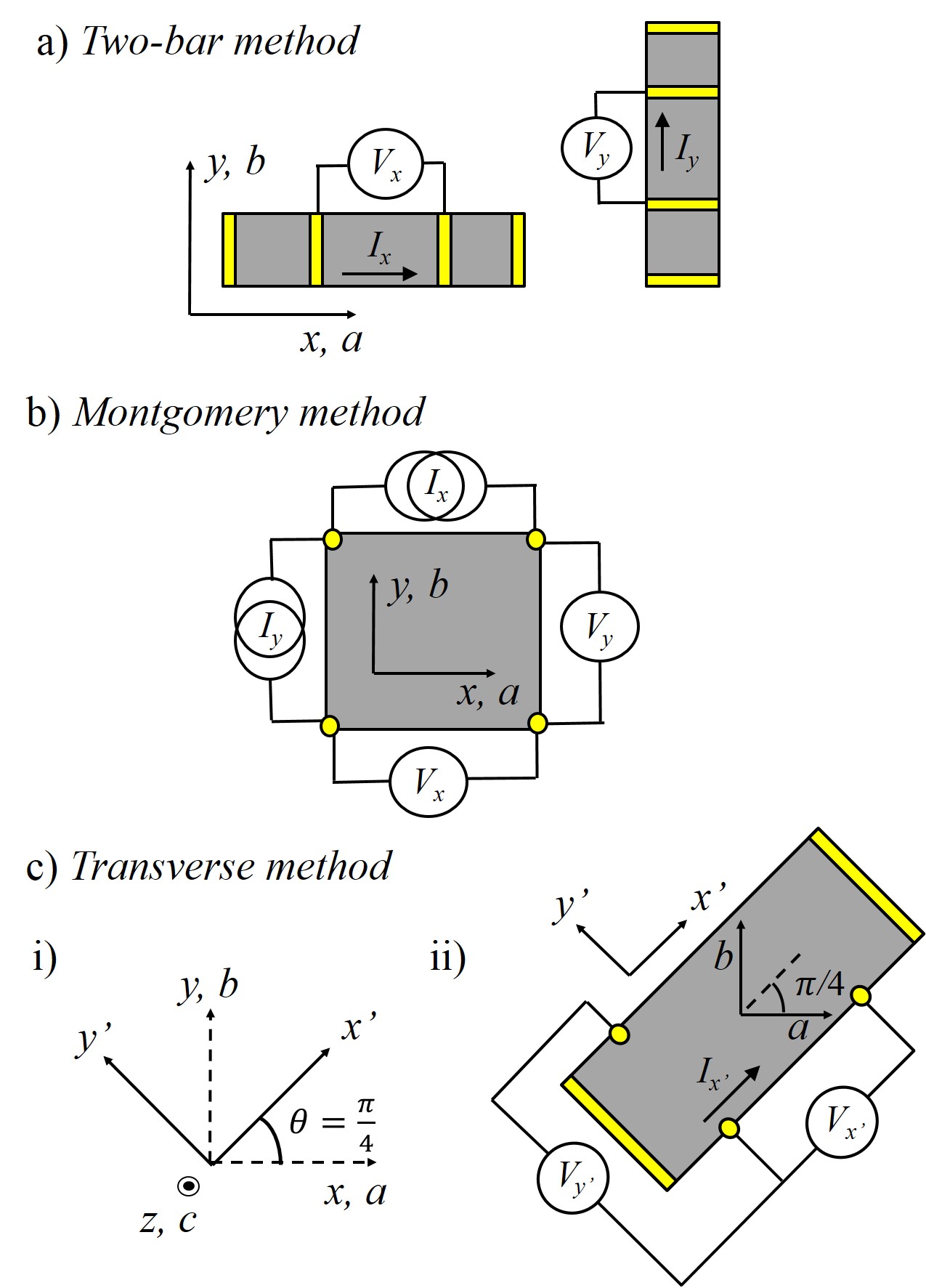}
\caption{Schematic diagrams illustrating different methods to determine the resistivity anisotropy of an orthorhombic material. a) The conventional two-bar method where a separate crystal is used for each component $\rho_{ii}$ to be measured. Any inequivalence between the two measurements or crystals results in admixing of the average resistivity $\nicefrac{1}{2}(\rho_{ii}+\rho_{jj})$ into the inferred resistivity anisotropy $(\rho_{ii}^m-\rho_{jj}^m)$, which can affect conclusions drawn about the temperature dependence and magnitude of $(\rho_{ii}-\rho_{jj})$. b) The Montgomery method uses a single rectilinear sample to measure the resistivity anisotropy with contacts on the corners, currents sourced parallel to the relevant crystallographic direction, and voltage measured across the opposite two corners. This contact geometry produces highly non-linear isopotentials in the sample that typically reduce the magnitude of the measured voltage by an order of magnitude or more relative to the bar method. Geometric factors also non-trivially mix the isotropic resistivity with the inferred resistivity anisotropy. c\,i) To motivate the new transverse method described in this paper, we consider a Cartesian coordinate system in which the measurement basis is rotated by an angle $\theta=\frac{\pi}{4}$ about the out-of-plane axis (shown here for the case of a rotation about $z$, $c$ for an $a$,$b$ plane anisotropy measurement) to produce a new basis $x'$, $y'$, $z'$ that is no longer aligned to the crystallographic axes $a$, $b$, $c$. As described in the main text, the absence of mirror planes $\sigma_{x'}$ and $\sigma_{y'}$ in this new basis results in finite off-diagonal terms in the resistivity tensor, the values of which are directly proportional to the resistivity anisotropy. c\,ii) The transverse method presented here uses a 5-point contact geometry as illustrated to measure the resistivity anisotropy in a single crystal that has been cut into a bar oriented along the diagonal of the measurement plane (the (110) direction for a measurement of $a$, $b$ plane anisotropy is the example shown here). This geometry does not reduce the magnitude of the voltage signal, and also measures the resistivity anisotropy directly via the transverse contacts (voltage $V_{y'}$ in the illustration). As discussed in the text, the transverse method is far less susceptible to admixing of the average resistivity into the resistivity anisotropy.}
\label{fig:contact_geometries}
\end{figure}

\subsubsection{\label{sec:2bar_method}Two-bar method}

The form of $\boldsymbol{\rho}$ is highly suggestive that the best way to measure $(\rho_{ii}-\rho_{jj})$ is with a current passed, and voltage measured, parallel to the relevant crystallographic axis. Hence the conventional two-bar method, illustrated in Figure \ref{fig:contact_geometries}a, whereby separate crystals are required to measure each component $\rho_{ii}$ and $\rho_{jj}$. Current and voltage contacts are ideally placed (respectively) on the ends and in the middle of the bar so as to short out all axes that are not being measured; some separation between current and voltage contacts is also desirable so as to ensure homogeneous current density in the case of uneven contact resistance.

\subsubsection{\label{sec:montgomery_method}Montgomery method}

An alternative method to separately determine individual terms in the resistivity tensor was deduced from the earlier work of van der Pauw by Montgomery for anisotropic materials\cite{vanderpauw1958method,montgomery1971method}. As shown in Figure \ref{fig:contact_geometries}b the Montgomery method uses contacts on the corners of a rectilinear sample, through which current is sourced parallel to either planar direction and voltage measured on the parallel opposite pair of contacts. Provided the sample is close to rectangular with edges well aligned to the crystallographic axes, and its dimensions well known, the measured voltages can be transformed to obtain the resistivity. The measured resistances are used to define the effective dimensions of the sample as if it were isotropic, the ``isotropic equivalent solid'' (from a measurement perspective, a square of anisotropic material can be equivalent to a rectangle of isotropic material). The intrinsic values of $\rho_{ii}$ and $\rho_{jj}$ are then calculated via a transformation involving the effective and real dimensions and the measured resistances\cite{montgomery1971method}.

\subsection{Sources of uncertainty for conventional methods}

There is a distinction between how systematic errors affect the absolute value of a single resistivity measurement and the determination of the resistivity anisotropy that varies between methods. The focus here is on minimising the error in $(\rho_{ii}-\rho_{jj})$, as well as considering the errors in $\rho_{ii}$ and $\rho_{jj}$ individually. The following discussion stresses effects that admix the resistivity anisotropy $(\rho_{ii} - \rho_{jj})$ and the average resistivity $\frac{\rho_{ii}+\rho_{jj}}{2}$, noting that these two quantities can have very different temperature dependencies. In particular, as we explain in greater detail below, any technique that aims to measure $(\rho_{ii}-\rho_{jj})$ must minimise admixture of $\frac{\rho_{ii}+\rho_{jj}}{2}$.

\subsubsection{\label{sec:2bar_errors}Two-bar method}
In the bar method the resistivity is derived from the measured resistance $R_{ii}^m$ by geometric factors:
\begin{equation}
\rho_{ii}^m=R_{ii}^m\frac{A^m}{l^m},
\end{equation}
where $A$ and $l$ are the cross-sectional area of the crystal and the voltage contact separation respectively, with the superscript $m$ indicating a measured value (as opposed to the intrinsic, error-free values). In principle each of these values has an error associated with its measurement, although the associated uncertainty in $R_{ii}^m$ is generally negligible in comparison to geometric errors and thus omitted from this discussion. $R_{ii}^m$ is however potentially sensitive to crystal misalignment: for a misaligned crystal,

\begin{equation}
R_{ii}^m=[\rho_{ii}(1-\cos^2(\theta))+\rho_{jj}\sin^2(\theta)]\frac{l}{A},
\end{equation}

where the misalignment $\theta$ is assumed for simplicity to be solely within the measurement plane, as described by Equation \ref{eqn:matrix_rot}. This is generally a reasonable assumption in layered materials. Ideally $\theta=0$, $l^m = l$ and $A^m = A$, but in any real measurement $\theta = 0+\Delta\theta$, $l^m = l+\Delta l$ and $A^m = A+\Delta A$. Since misalignment enters $R_{ii}^m$ as $\theta^2$ for small $\theta$ it can be treated as a weak perturbation in most materials and we neglect it here, focusing instead on the more significant geometric factors. In particular, when considering the contribution of geometric errors,

\begin{equation}
\rho_{ii}^m=\rho_{ii}\frac{A^m}{l^m}\frac{l}{A}\approx\rho_{ii}\bigg(1+\frac{\Delta A}{A}-\frac{\Delta l}{l}\bigg),
\label{eqn:barerrors}
\end{equation}

the error in $\rho_{ii}$ is found to be linear in $\Delta A$ and $\Delta l$ and so these errors will dominate. It should be noted however that in the special case of extremely anisotropic materials ($1000\rho_{ii} \approx \rho_{jj}$) contact or crystal misalignment can be the leading error\cite{mercure2012lmo,hussey2002pr124}. $\Delta A$ and $\Delta l$ can both be large in a single measurement, but only appear as multiplicative factors and thus do not affect the temperature dependence of any given component $\rho_{ii}$. Taken in isolation, these errors do not affect any physical conclusions drawn from $\rho_{ii}$ other than the magnitude. However this situation changes when considering the difference between two componenents $(\rho_{ii}-\rho_{jj})$. To illustrate this point we can characterise each measurement as,

\begin{equation}
\begin{split}
\rho_{ii}^m&=\rho_{ii}(1+\Delta_{ii})\\
\rho_{jj}^m&=\rho_{jj}(1+\Delta_{jj}),
\end{split}
\label{eqn:err_prop}
\end{equation}

and then propagate the measurement errors when finding $\rho_{ii}-\rho_{jj}$,

\begin{equation}
\begin{split}
\rho_{ii}^m-\rho_{jj}^m=&\frac{1}{2}(2+\Delta_{ii}+\Delta_{jj})(\rho_{ii}-\rho_{jj})\\
+&\frac{1}{2}(\Delta_{ii}-\Delta_{jj})(\rho_{ii}+\rho_{jj}),
\end{split}
\label{eqn:errorprop}
\end{equation}

thus showing that terms with different temperature dependences (i.e. $(\rho_{ii}+\rho_{jj})$ and $(\rho_{ii}-\rho_{jj})$) can become admixed. This could fundamentally change the conclusions drawn from an experiment. For example, changes in the average resistivity $(\rho_{ii}+\rho_{jj})$ at a phase transition would lead to an apparent change in the measured anisotropic resistivity $(\rho_{ii}^m-\rho_{jj}^m)$ if $\rho_{ii}$ and $\rho_{jj}$ are not measured precisely. One would then be led to a potentially erroneous conclusion that the phase transition breaks C$_4$ symmetry when in fact it might not.

Whilst the geometric effects are typically the leading error, there are other experimental factors that can lead to an inaccurate determination of $\rho_{ii}-\rho_{jj}$ via the two-bar method. Errors in the real current density can arise through the sample shape not being strictly oblong, and also through non-ideal current paths, either due to poor contact placement or sample homogeneity issues. In macroscopic samples the former can be assessed optically and corrected to within acceptable error by cleaving, polishing, or some other sample manipulation, and thus is unlikely to form a dominant error. In thin-films however a small change in the thickness of the sample locally can have a significant effect. As illustrated in Figure \ref{fig:contact_geometries}a the current contacts should cover the end of the bar, but also the out-of-plane axis so as to `short out' the axes that are not to be measured. If this is not performed correctly then the current density will be uneven within some characteristic length scale of the contacts, and the current path may not be strictly parallel to the axis to be measured, thus contaminating the signal in an uncontrolled fashion.

Even in a perfectly performed measurement, the two-bar method relies on both samples being of identical composition and purity. For example, when calculating $(\rho_{ii}-\rho_{jj})$ a slight compositional change could lead to a single phase transition appearing as two if it occurs at a slightly different temperature in each sample. At low temperature the purity becomes very important as the residual resistivity, which is approximately proportional to the impurity density, dominates the signal. Thus samples of different purity would appear to indicate a non-zero $(\rho_{ii}-\rho_{jj})$ even in an isotropic system.

Finally we note that when performing a resistivity measurement it is assumed that the sample temperature is well represented by a nearby thermometer; however, in practice this is often not the case due to thermal gradients. Hence the temperature can vary between samples depending on exactly how they are attached to the measurement stage, proximity to the thermometer, or local thermal fluctuations e.g. from uneven gas flow in a flow cryostat. This can produce an uneven temperature error between the samples, which has an acute effect on $(\rho_{ii}-\rho_{jj})$ in the case that $\nicefrac{d\rho_{ii}}{dT}$ or $\nicefrac{d\rho_{jj}}{dT}$ is large.

The key point from this discussion of the two-bar method is that $(\rho_{ii}-\rho_{jj})$ is extremely sensitive to inequivalences in the measurement environment, composition and contact placement of the two samples. This can generally be characterised in terms of admixing between average and anisotropic components of the resistivity. Errors in $\rho_{ii}$ or $\rho_{jj}$ do not just cause an error in the magnitude of $(\rho_{ii}-\rho_{jj})$ but can also give the wrong temperature dependence and the appearance of a significant finite value even for cases where intrinsically $(\rho_{ii}-\rho_{jj})=0$ (the case for tetragonal materials). This is particularly acute for the case $(\rho_{ii}-\rho_{jj}) \ll \nicefrac{1}{2}(\rho_{ii}+\rho_{jj})$ (small anisotropies) where the magnitude of the admixing can dwarf the real anisotropy.

\subsubsection{\label{sec:montgomery_errors}Montgomery method}

The Montgomery method allows for measurement of $(\rho_{ii}-\rho_{jj})$ in a single sample, with $\rho_{ii}$ and $\rho_{jj}$ in principle measured simultaneously if care is taken over instrumentation\cite{kuo2016ubiquitous}. This precludes some of the errors that arise in the two-bar method but at the expense of other effects arising from the contact geometry (see Figure \ref{fig:contact_geometries}b).
This is because the Montgomery method is derived via a number of assumptions that are not always met experimentally. 

For the Montgomery method to be valid the sample must be square or rectangular in the plane and of constant thickness. It has been shown that it is possible to generalise this situation slightly to samples that are parallelograms in the plane and thus only a single additional parameter is required to describe the geometry, but each additional parameter increases the complexity of the analysis, and it may not be possible to solve all situations analytically\cite{borup2012parallelogram}. The error induced by deviations from the expected relative angle of the sample edges, $\Delta\phi$ is divergent as angle increases and contributes equally and oppositely to $\rho_{ii}^m$ and $\rho_{jj}^m$. For example, an error of $\Delta\phi=$2\degree\,gives $\nicefrac{\Delta_{ii}}{\rho_{ii}}=-\nicefrac{\Delta_{jj}}{\rho_{jj}}=0.01$ with this value increasing to 0.04 for 4\degree. Again considering Equation \ref{eqn:errorprop}, this admixes the average resistivity $\frac{1}{2}(\rho_{ii}+\rho_{jj})$ proportionately into the measured $(\rho_{ii}^m-\rho_{jj}^m)$.

The Montgomery method is acutely sensitive to current paths, and as such it is crucial that the out-of-plane thickness is constant and the sample homogeneous. The magnitude of the thickness is also a non-trivial consideration: the measured voltages have a non-linear relationship with sample thickness as the thickness of the``equivalent isotropic solid" becomes of the same order as the in-plane dimensions\cite{montgomery1971method}, which can be an issue even in thin samples of highly 2-D materials. The error induced by non-uniform sample thickness is typically of order the proportional change in thickness but can be greater, however the exact topology of the sample and contact positioning is important and so this error is difficult to treat generally. Sample inhomogeneity is essentially a local distortion of the``equivalent isotropic solid" and so should be considered similarly.

Finally, finite contact size is a non-trivial error in the Montgomery technique for the case where the contacts are not negligibly small relative to the sample dimensions. In a bar measurement, using an effective contact centre is a simple solution; however, the same solution applied to the Montgomery method effectively breaks the assumption that the contacts are on the edges of the sample. Again, the induced errors are non-linear and difficult to quantify, but must alter the topography of the isopotentials in the sample thus affecting the measured voltage in an uncontrolled way.

Finally, an important experimental consideration with the Montgomery method is the reduction in magnitude of the measured voltage due to non-parallel equipotentials produced by the contact geometry (relative to an equivalently sized bar). Typically this reduction can be an order of magnitude in samples with favourable aspect ratios, but becomes far higher as anisotropy increases\cite{bierwagen2004vdpcalc}. 

\subsection{Measurement of resistivity anisotropy by the transverse method}

As an alternative to the two-bar and Montgomery methods, we note that the resistivity anisotropy can be accessed directly if we relax the constraint that $a\parallel x$, $b\parallel y$ and $c\parallel z$ under which $\boldsymbol{\rho}$ is conventionally described, and rotate the Cartesian basis (in which the vectors $J_j$ and $E_i$ are defined) relative to the crystallographic axes about the out-of-plane axis by an angle $\theta$ as illustrated in Figure \ref{fig:contact_geometries}c)i. Assuming anisotropy is to be measured in the $a$, $b$ plane the conventional resistivity tensor is thus rotated to obtain,

\begin{equation}
R _{z,c}(\theta) \boldsymbol{\rho} R_{z,c}^T (\theta)=
\label{eqn:matrix_rot}
\end{equation}
\begin{equation*}
\begin{pmatrix}
\rho_{a}\cos^2(\theta)+\rho_{b}\sin^2(\theta)&(\rho_{a}-\rho_{b})\cos(\theta)\sin(\theta)&0\\
(\rho_{a}-\rho_{b})\cos(\theta)\sin(\theta)&\rho_{a}\sin^2(\theta)+\rho_{b}\cos^2(\theta)&0\\
0&0&\rho_{c}\\
\end{pmatrix},
\end{equation*}
where $R_\alpha$ is the rotation operator about axis $\alpha$, taken here to be the $z$, $c$ axis. By setting $\theta = \frac{\pi}{4}$ we obtain,

\begin{equation}
\boldsymbol{\rho}' = \frac{1}{2}
\begin{pmatrix}
\rho_{a}+\rho_{b}&\rho_{a}-\rho_{b}&0\\
\rho_{a}-\rho_{b}&\rho_{a}+\rho_{b}&0\\
0&0&2\rho_{c}\\
\end{pmatrix}
,
\label{eqn:rot_rho}
\end{equation}

which contains off-diagonal components $\rho'_{x'y'}$ and $\rho'_{y'x'}$ that directly give the in-plane resistivity anisotropy $\frac{\rho_{a}-\rho_{b}}{2}$. Herein primed notation indicates the rotated Cartesian basis, thus describing an experiment where $J_{x'}$ represents a current applied at an angle $\theta=\nicefrac{\pi}{4}$ to the $a$ axis and so on. The diagonal components $\rho'_{x'x'}$ and $\rho'_{y'y'}$ give the mean of the in-plane resistivities, $\frac{\rho_{a}+\rho_{b}}{2}$. $\rho_{a}$ and $\rho_{b}$ can therefore be deduced by combining diagonal and off-diagonal components:
\begin{equation}
\begin{gathered}
\rho_{a} = \rho'_{x'x'} + \rho'_{x'y'}\\
\rho_{b} = \rho'_{x'x'}-\rho'_{x'y'}.
\end{gathered}
\label{eqn:rhoa_rhob}
\end{equation}
This derivation can be trivially repeated for other measurement planes (indeed in section \ref{sec:ErTe3} the measurement is demonstrated in the $a-c$ plane for ErTe$_3$). The experimental configuration shown in Figure \ref{fig:contact_geometries}c)ii allows the simultaneous measurement of both $\rho'_{x'x'}$ and $\rho'_{x'y'}$ in a single sample. A current is passed along the crystallographic (1 1 0) direction, and voltage measured both parallel and perpendicular to this current. The measured resistances are converted to resistivities in the usual fashion\cite{normalisationcomment}.

The transverse voltage is allowed due to the lack of a mirror plane perpendicular to the $y'$ direction; these mirror planes are necessarily absent in the presence of finite resistivity anisotropy as the resistivity tensor must obey the symmetries of the point group, and no orthorhombic crystal can have diagonal mirror planes. It is important to stress that this measurement scheme is not to be confused with a Hall effect measurement despite the similarities in contact geometry; Hall resistivities are odd under time-reversal and thus zero in the absence of magnetic field or magnetic order in contrast to the present result in Equation \ref{eqn:rot_rho} which is even under time-reversal. An alternative (but fundamentally equivalent) explanation for the origin of the transverse resistivity in this rotated reference frame is given in Appendix A.

\subsection{\label{sec:transerrors}Sources of uncertainty for the transverse method}

\begin{figure}
\includegraphics[width=0.5\columnwidth]{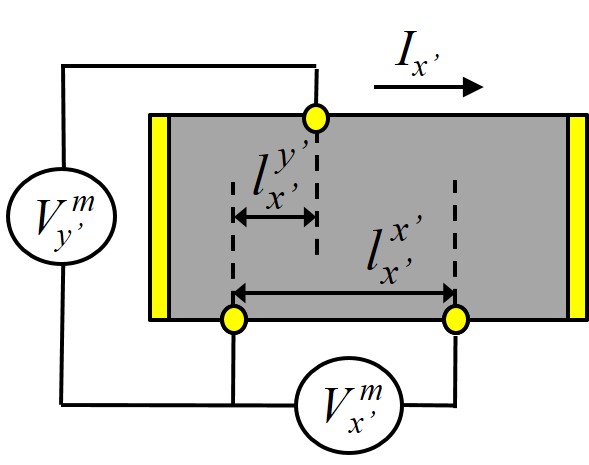}
\caption{Unintentional misalignment of transverse contacts in the transverse method causes the measured transverse voltage $V^m_y$ to be contaminated by a longitudinal voltage such that $V^m_{y'}=V_{y'}+(\nicefrac{l^{y'}_{x'}}{l^{x'}_{x'}})V_{x'}$ where $l^{x'}_{x'}$ and $l^{y'}_{x'}$ are the separation in the $x'$ direction of the longitudinal and transverse contacts respectively, as illustrated in the figure ($l^{y'}_{x'}$ is exaggerated for clarity). This accidental misalignment can often be corrected either exactly or approximately when considering the symmetry of the crystal, as described in the main text.}
\label{fig:contact_misalignment}
\end{figure}

An important experimental concern with the transverse method is that without exceptional care the transverse contacts (nominally measuring $V_{y'}$ in Figure \ref{fig:contact_geometries}c)ii) will never be truly perpendicular to the current as illustrated in Figure \ref{fig:contact_misalignment}. The accidental offset of these contacts leads to a contamination of the transverse voltage characterised by 
\begin{equation}
V^m_{y'} = V_{y'} + \frac{l^{y'}_{x'}}{l^{x'}_{x'}}V^m_{x'},
\label{eqn:contact_offset}
\end{equation}
where $V^m_{y'}$ is the voltage measured across the real, misaligned, transverse contacts, $V_{y'}$ the intrinsic transverse voltage, $V^m_{x'}$ the measured longitudinal voltage, $l^{x'}_{x'}$ the measured spacing between the longitudinal voltage contacts, and $l^{y'}_{x'}$ the accidental offset in the $x$ direction of the transverse voltage contacts. If $V_{y'}$ can be assumed to be zero in some regime due to the resistivity being isotropic then Equation \ref{eqn:contact_offset} allows the determination of $\nicefrac{l^{y'}_{x'}}{l^{x'}_{x'}}$ in this regime. As $\nicefrac{l^{y'}_{x'}}{l^{x'}_{x'}}$ is a temperature independent geometric factor that is constant throughout the measurement, this allows the contamination signal to be subtracted across the whole range of measurement (as $V^m_{x'}$ is also measured). This is not a trivial assumption, but it can be explicitly tested by checking whether $\nicefrac{V^m_{y'}}{V^m_{x'}}$ is constant in the isotropic regime. This condition is perfectly satisfied in the case of a tetragonal to orthorhombic distortion; however, the applicability to the other case discussed in section \ref{sec:introduction} of an orthorhombic material that gains additional anisotropy depends on the specifics of that material. In section \ref{sec:ErTe3} we argue that the assumption is valid for ErTe$_3$ and its application is demonstrated in section \ref{sec:results}, but this is not a general statement.

The transverse method does not avert errors in sample geometry and finite contact size entirely, but it does eliminate the admixing effects described by Equation \ref{eqn:errorprop} when determining $(\rho_{ii}-\rho_{jj})$. The influence of geometric measurement error is now described by,

\begin{equation}
\rho_{x'y'}^{\prime m}=\rho'_{x'y'}\bigg(1+\frac{\Delta A}{A}-\frac{\Delta l_{y'}^{y'}}{l_{y'}^{y'}}\bigg)=\rho'_{x'y'}(1+\Delta_{x'y'}),
\label{eqn:geometric_err}
\end{equation}

where $l_{y'}^{y'}$ is the separation in the $y'$ direction of the transverse contacts. Provided the longitudinal contamination signal is correctly subtracted as described above, this shows that geometric errors only manifest as a prefactor to the resistivity anisotropy and do not affect its temperature dependence by admixing the average resistivity $\nicefrac{1}{2}(\rho_{ii}+\rho_{jj})$, in contrast to the two-bar method and Montgomery method. This is the principal advantage of the transverse method.

Angular alignment errors are again effectively derived from Equation \ref{eqn:matrix_rot} and contribute to the measurement as,

\begin{equation}
\rho_{x'y'}^{\prime m}=(\rho_a-\rho_b)\cos\bigg(\frac{\pi}{4}+\Delta\theta\bigg)\sin\bigg(\frac{\pi}{4}+\Delta\theta\bigg),
\end{equation}

which is linear in small $\Delta\theta$, but introduces no admixture of $\rho'_{x'x'}$ (inspection of the component $\rho'_{x'y'}$ in the transformed resistivity tensor in Equation \ref{eqn:matrix_rot} makes it clear that angular error does not admix $(\rho_{a}+\rho_{b})$). As with the geometric errors this gives a prefactor to the resistivity anisotropy without altering the apparent temperature dependence.

For the purpose of geometric error propagation, the measurement of the average resistivity $\rho'_{x'x'}$ can be considered like a single bar method, thus the associated errors are analogous to those described in Equation \ref{eqn:barerrors}. From Equation \ref{eqn:matrix_rot}, misalignment errors give,

\begin{equation}
\rho_{x'x'}^{\prime m}=\rho_a\cos^2\bigg(\frac{\pi}{4}+\Delta\theta\bigg)+\rho_b\sin^2\bigg(\frac{\pi}{4}+\Delta\theta\bigg),
\end{equation}

and it can be seen that $\rho_{x'x'}^{\prime m}$ becomes weighted towards either $\rho_{ii}$ or $\rho_{jj}$ with finite $\Delta\theta$. As $\nicefrac{d\cos^2(\theta)}{d\theta}$ and $\nicefrac{d\sin^2(\theta)}{d\theta}$ are equal and large in magnitude but opposite in sign around $\theta=\frac{\pi}{4}$ this effect can be appreciable if there is significant anisotropy but largely cancels out in more weakly anisotropic systems. 

The error propagation when determining $\rho_a$ and $\rho_b$ individually via Equation \ref{eqn:rhoa_rhob} using the transverse method is directly analogous to that when determining $(\rho_a-\rho_b)$ via the two-bar method. With reference to Equations \ref{eqn:err_prop} and \ref{eqn:geometric_err}, and considering just geometric errors (i.e. neglecting angular misalignment), the values measured by the transverse technique can be written as

\begin{equation}
\begin{split}
(\rho_a - \rho_b)^m&=(\rho_a-\rho_b)(1+\Delta_{x'y'})\\
(\rho_a + \rho_b)^m&=(\rho_a + \rho_b)(1+\Delta_{x'x'}),
\end{split}
\end{equation}

and thus combine to give

\begin{equation}
\begin{split}
\rho_{a}^m&=\rho_a\bigg(1-\frac{\Delta_{x'y'}}{2}+\frac{\Delta_{x'x'}}{2}\bigg)\\&+\rho_b\bigg(\frac{\Delta_{x'x'}}{2} -\frac{\Delta_{x'y'}}{2}\bigg)
\end{split}
\end{equation}
\begin{equation}
\begin{split}
\rho_b^m&=\rho_b\bigg(1+\frac{\Delta_{x'x'}}{2}+\frac{\Delta_{x'y'}}{2}\bigg)\\&+\rho_a\bigg(\frac{\Delta_{x'x'}}{2}-\frac{\Delta_{x'y'}}{2}\bigg).
\end{split}
\end{equation}

Evidently $\rho_a$ can suffer an admixture of $\rho_b$ and vice versa when determined individually via the transverse method, in contrast to the two-bar method where this mixing does not occur. This is analogous to the admixing of $(\rho_a-\rho_b)$ and $(\rho_a + \rho_b)$ in the two-bar method, which does not occur in the transverse method.

In summary, the measurement of $(\rho_{ii}-\rho_{jj})$ by the transverse method is very robust against admixture from the average resistivity, giving a significant improvement on the two-bar and Montgomery methods provided that the contamination signal due to accidental contact offset can be subtracted or minimised. Furthermore, the direct measurement of $(\rho_{ii}-\rho_{jj})$ vastly improves signal to noise by effectively removing the isotropic `background' signal. It is noted that $\frac{\rho_{ii}+\rho_{jj}}{2}$ can become unevenly weighted when measured via the longitudinal contacts due to angular misalignment, but this effect is unlikely to be larger than the contribution of geometric errors in the two-bar method. Measurements of the individual components of the resistivity tensor $\rho_a^m$ and $\rho_b^m$ suffer the effects of admixing between $\rho'_{x'x'}$ and $\rho'_{x'y'}$, and so the transverse method may not offer an improvement over a single bar when measuring a single component. However when comparing two components, there is clearly a significant advantage to the transverse method over the two-bar and Montgomery methods, which is the principal message of this paper.

\section{\label{sec:ErTe3}Demonstration of the transverse method: measurement of resistivity anisotropy in ErTe$_3$}

\subsection{Resistivity anisotropy in ErTe$_3$}

\begin{figure}
\includegraphics[width=\columnwidth]{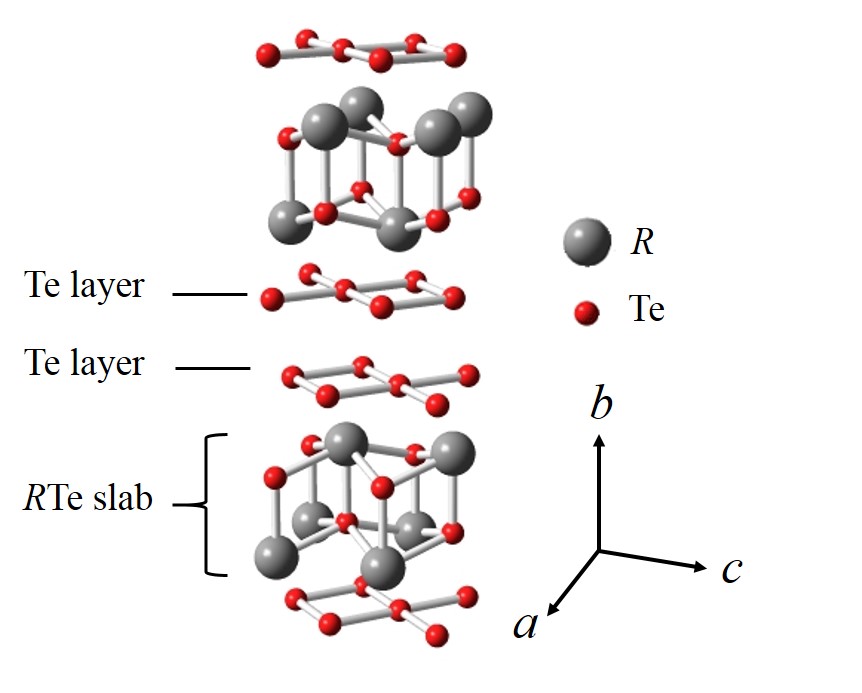}
\caption{The structure of $R$Te$_3$ consists of conducting Te bilayers in the $a-c$ plane that are sandwiched between insulating $R$Te slabs. The well separated conducting planes produce a highly two-dimensional Fermi surface that is well described by a simple tight-binding model\cite{laverock2005rte3fs}. As the unit cell is stacked in the $b$ direction each subsequent unit is offset by half a unit cell in the $a$ direction giving a glide-plane symmetry to the structure. This glide plane is the source of the slight orthorhombicity in the material.} 
\label{fig:structure+fs}
\end{figure}

In order to demonstrate the efficacy of the transverse technique described in the previous section we have applied the technique to the layered rare-earth tritelluride ErTe$_3$. The rare-earth tritellurides form for $R$=Y, La-Sm, Gd-Yb\cite{ru2008chempress}. At high temperature they have the NdTe$_3$ structure type (Cmcm) consisting of RTe blocks separating almost square bilayer Te planes stacked vertically as illustrated in Figure \ref{fig:structure+fs}. The single layer compound $R$Te$_2$ has similar motifs ($R$Te block with a single Te layer) and is tetragonal at high temperature. However, $R$Te$_3$ has a glide plane that causes the material to be very slightly orthorhombic ($a\approx 0.9995c$)\cite{ru2008chempress}. Upon cooling, a uni-directional CDW forms along the $c$ direction for all $R$, with heavier $R$ (Tb-Yb) also forming a second CDW along the $a$ direction at lower temperatures. The calculated Fermi surface in the absence of CDW ordering is found to be essentially isotropic in the $a-c$ plane (reflecting the almost vanishingly small difference in the $a$ and $c$ lattice parameters), as well as highly two-dimensional, with almost no dispersion in the $b$-axis direction\cite{laverock2005rte3fs}. This is because the Fermi surface is almost entirely derived from Te $p_x$ and $p_z$ states in the Te square-net bilayers. Thus in the absence of CDW ordering the material is ``almost tetragonal'' in the context of electrical transport. The orthorhombicity only becomes a signficant factor very close to the CDW transition temperature - phonon frequencies soften in both the $a$ and $c$ directions above the CDW transition temperature $T_{c1}$\cite{maschek2015wave}, but only go to zero in the $c$ direction thus stabilising a mono-domain, uni-directional CDW rather than a bi-directional CDW or domains of perpendicular, uni-direcitonal CDWs. The resultant gapping of the Fermi surface induces significant anisotropy into the electrical transport, thus placing $R$Te$_3$ into the second category of material discussed in the introduction: orthorhombic materials that gain additional anisotropy\cite{sinchenko2014montgomery,moore2010ErTe3}. A crucial advantage with $R$Te$_3$ over other examples in this class is that the process described in section \ref{sec:transerrors}, whereby longitudinal contamination of the transverse voltage can be subtracted, is applicable owing to the highly isotropic transport properties above $T_{c1}$. This combination of properties make $R$Te$_3$ the perfect material to demonstrate the efficacy of the transverse method, with the specific example of ErTe$_3$ selected for its convenient CDW ordering temperatures, $T_{c1}$=\SI{267}{\kelvin} (CDW ordering $\parallel c$)\cite{ru2008chempress} and $T_{c2}$=\SI{160}{\kelvin} (CDW ordering $\parallel a$)\cite{moore2010ErTe3}.

\subsection{\label{sec:experimental_methods}Experimental Methods}

\begin{figure}
\includegraphics[width=0.6\columnwidth]{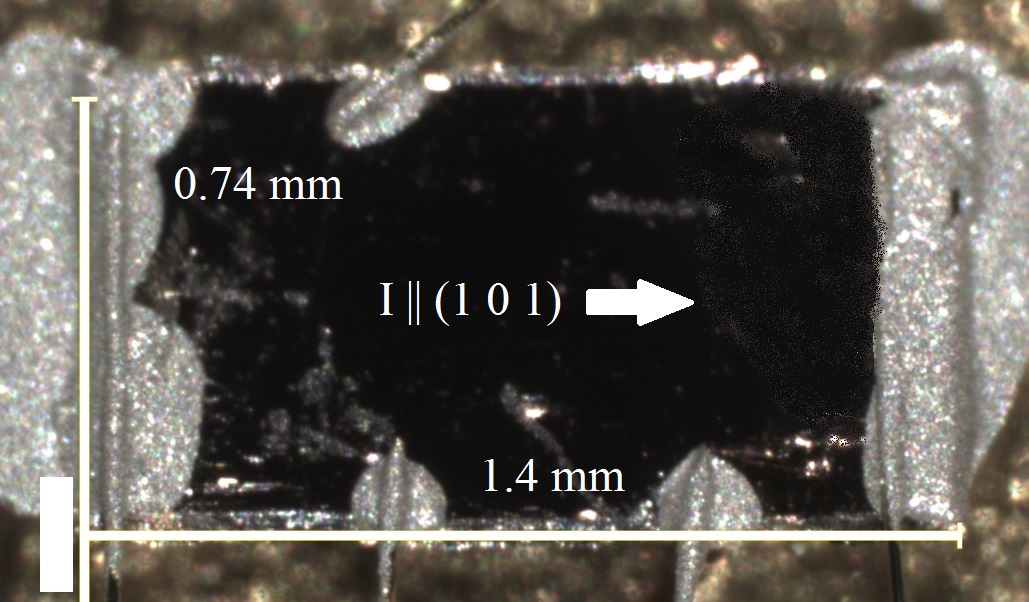}
\caption{An optical microscope image of the contacted sample illustrating the contact geometry and finite contact size. The real sample has a gold hue that is not well represented in this image.}
\label{fig:sample}
\end{figure}
The experiment was performed using a Quantum Design PPMS temperature controller, voltages were measured via two phase-locked Stanford Research Systems SR830 lock-in amplifiers with current sourced from the reference lock-in's voltage output via a \SI{4.5}{\kilo\ohm} pre-resistor to give a \SI{1}{\milli\ampere} current. A nominal gain of 1000 was achieved by combination of a Princeton research model 1900 transformer (100$\times$ step-up) and a Stanford Research SR560 pre-amplifier (10$\times$ gain) on each voltage channel, and then calibrated. Single crystals of ErTe$_3$ were grown via a self-flux method as described elsewhere\cite{ru2006growth}, and aligned by x-ray diffraction. The sample was cleaved in the $a-c$ plane and then cut to produce a bar in the (1 0 1) direction using a scalpel blade, with errors minimised to less than 5\degree\,by measuring the angle via optical microscope in relation to edges of the as-grown crystal that form in the (1 0 0) and (0 0 1) directions. The cut crystal was \SI{1.4}{\milli\meter} $\times$ \SI{0.74}{\milli\meter} $\times$ \SI{50}{\micro\meter} in $x, z$ and $y$ dimensions. Electrical contacts were made by sputtering gold pads through a mask and then attaching \SI{25}{\micro\meter} gold wires with Dupont 4929N silver paste with the contact geometry illustrated in Figure \ref{fig:contact_geometries}d. Care was taken to ensure the current contacts fully covered the end of the bar and shorted the out-of-plane axis. The contacted crystal is shown in Figure \ref{fig:sample}. Voltage contact separation was measured to the centre of the contacts with the longitudinal contacts separated by \SI{0.49}{\milli\meter} and the transverse contacts separated by \SI{0.64}{\milli\meter}. The contacts were 50 - \SI{100}{\micro\meter} in size with the offset between the transverse contacts significantly less than the contact size.

\subsection{\label{sec:results}Results}

\begin{figure}
\includegraphics[width=\columnwidth]{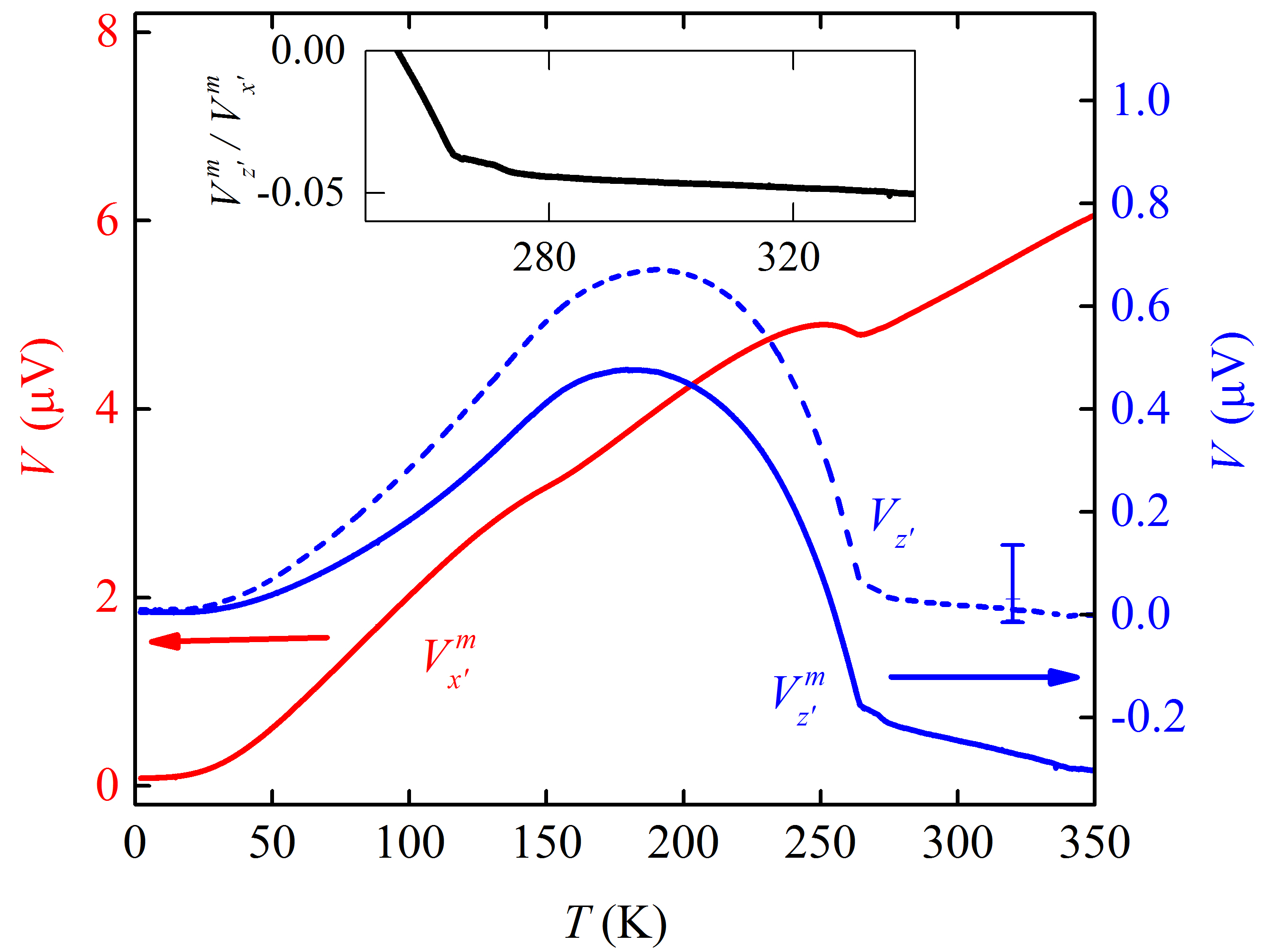}
\caption{\emph{Main:} raw voltage data $V^m_{x'}$ (red, left scale) and $V^m_{z'}$ (blue, right scale). Following correction for contact misalignment as described in the main text, $V_{z'}$ is shown as the dashed blue line, with the error bar indicating reasonable uncertainty in the offset of the effective point contacts as determined optically. It is assumed that $V_{x'}=V^m_{x'}$. \emph{Inset:} the ratio $\nicefrac{V^m_z}{V^m_x}$ is shown to be approximately constant above $T_{c1}$, indicating that the in-plane resistivity can be reasonably approximated as isotropic in this region according to Equation \ref{eqn:contact_offset}.}
\label{fig:raw_data}
\end{figure}

\begin{figure}
\includegraphics[width=\columnwidth]{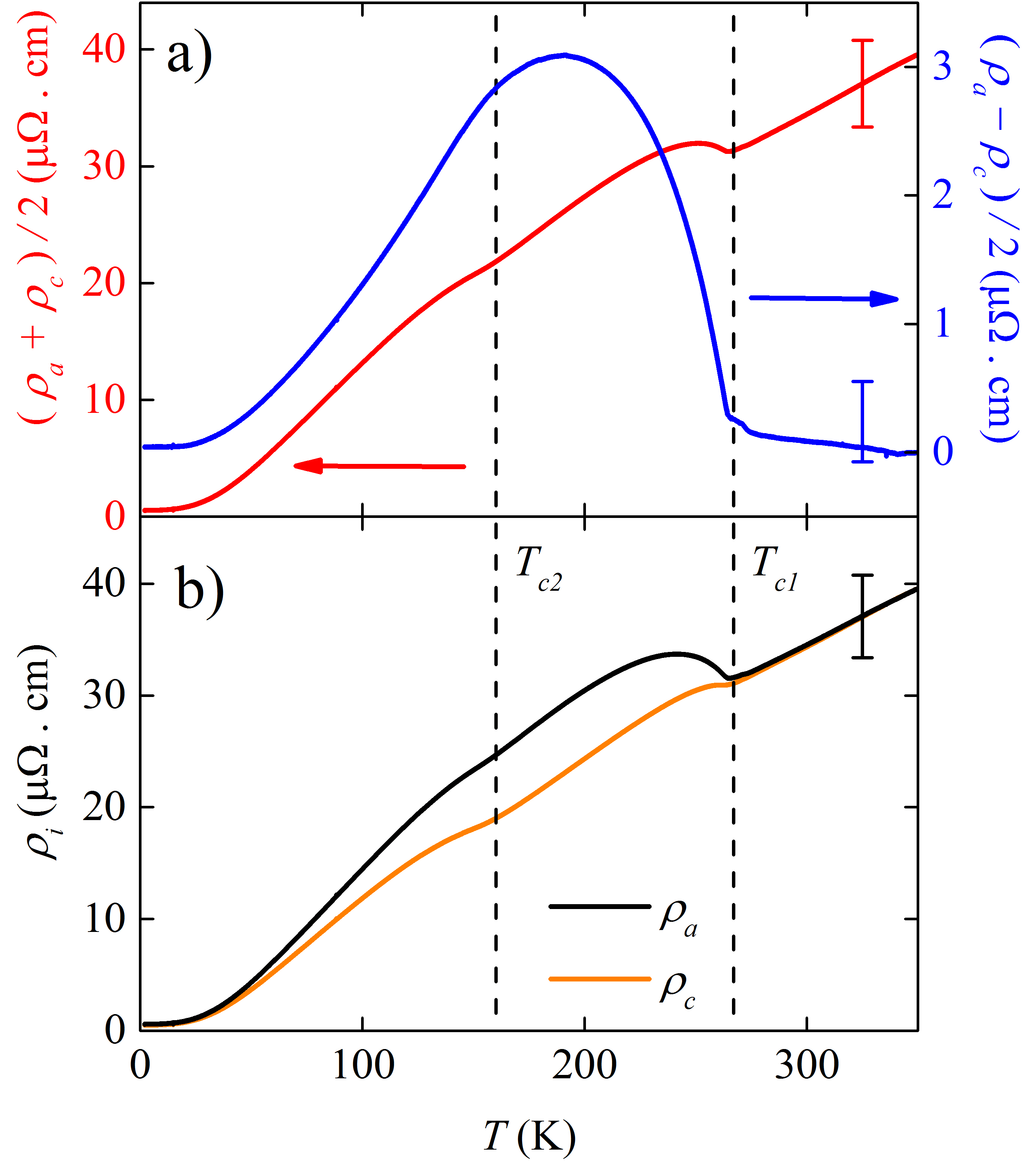}
\caption{a) Average resistivity $\frac{\rho_{a}+\rho_{c}}{2}$ (red, left scale) and resistivity anisotropy $\frac{\rho_{a}-\rho_{c}}{2}$ (blue, right scale). The error in $\frac{\rho_{a}+\rho_{c}}{2}$ is estimated from geometric uncertainties, with the dominant error in  $\frac{\rho_{a}-\rho_{c}}{2}$ coming from the correction of longitudinal contamination. b) The calculated values of $\rho_a$ and $\rho_c$ found via Equation \ref{eqn:rhoa_rhob}, the systematic error is dominated by the geometric error in $\frac{\rho_{a}+\rho_{c}}{2}$ and is necessarily the same in both $\rho_a$ and $\rho_c$.}
\label{fig:analysed_data}
\end{figure}

Figure \ref{fig:raw_data} shows the measured transverse (blue) and longitudinal (red) voltages ( $V_{z'}^m$ and $V^m_{x'}$ respectively). The inset shows that the ratio $\nicefrac{V^m_{z'}}{V^m_{x'}}$ is approximately constant above $T_{c1}$, which by reference to Equation \ref{eqn:contact_offset} is consistent with almost isotropic in-plane resistivity in the normal state and a contact offset of 17\,$\mu$m. The blue dashed line in Figure \ref{fig:raw_data} shows the transverse voltage corrected for this inferred offset, labelled $V_{z'}$. As the offset is a little smaller than the contact size, the error in this correction was estimated by the uncertainty in the position of the effective point contacts, determined optically, with the inferred value found to be within this range. Using these corrected values, the inferred values of $\frac{\rho_{a}-\rho_{c}}{2}$ and $\frac{\rho_{a}+\rho_{c}}{2}$ are shown in Figure \ref{fig:analysed_data}a. Both before and after the subtraction of the contamination signal, the onset of anisotropy below $T_{c1}$ dominates the transverse signal, highlighting the sensitivity of this technique to changes in anisotropy. The dominant error in $\frac{\rho_{a}+\rho_{c}}{2}$ is geometric uncertainty due to angular alignment, finite contact sizes, and a relatively thin sample, whereas in $\frac{\rho_{a}-\rho_{c}}{2}$ the dominant error is due to the uncertainty in the subtraction of the longitudinal contamination. Note that if ErTe$_3$ were truly tetragonal above $T_{c1}$ then there would be almost no uncertainty in this subtraction.

The resistivities $\rho_{a}$ and $\rho_{c}$ were derived via Equation \ref{eqn:rhoa_rhob} and are plotted in Figure \ref{fig:analysed_data}b. Note that no interpolation or fitting was required to add and subtract the data owing to the simultaneous, single crystal measurement. The values obtained are consistent with those found by conventional methods and published elsewhere\cite{sinchenko2014montgomery,ru2008chempress}. The systematic errors are dominated by geometric errors in $\frac{\rho_{a}+\rho_{c}}{2}$, that crucially must be identical in the determination of both $\rho_a$ and $\rho_c$. 

\begin{figure}
\includegraphics[width=0.88\columnwidth, left]{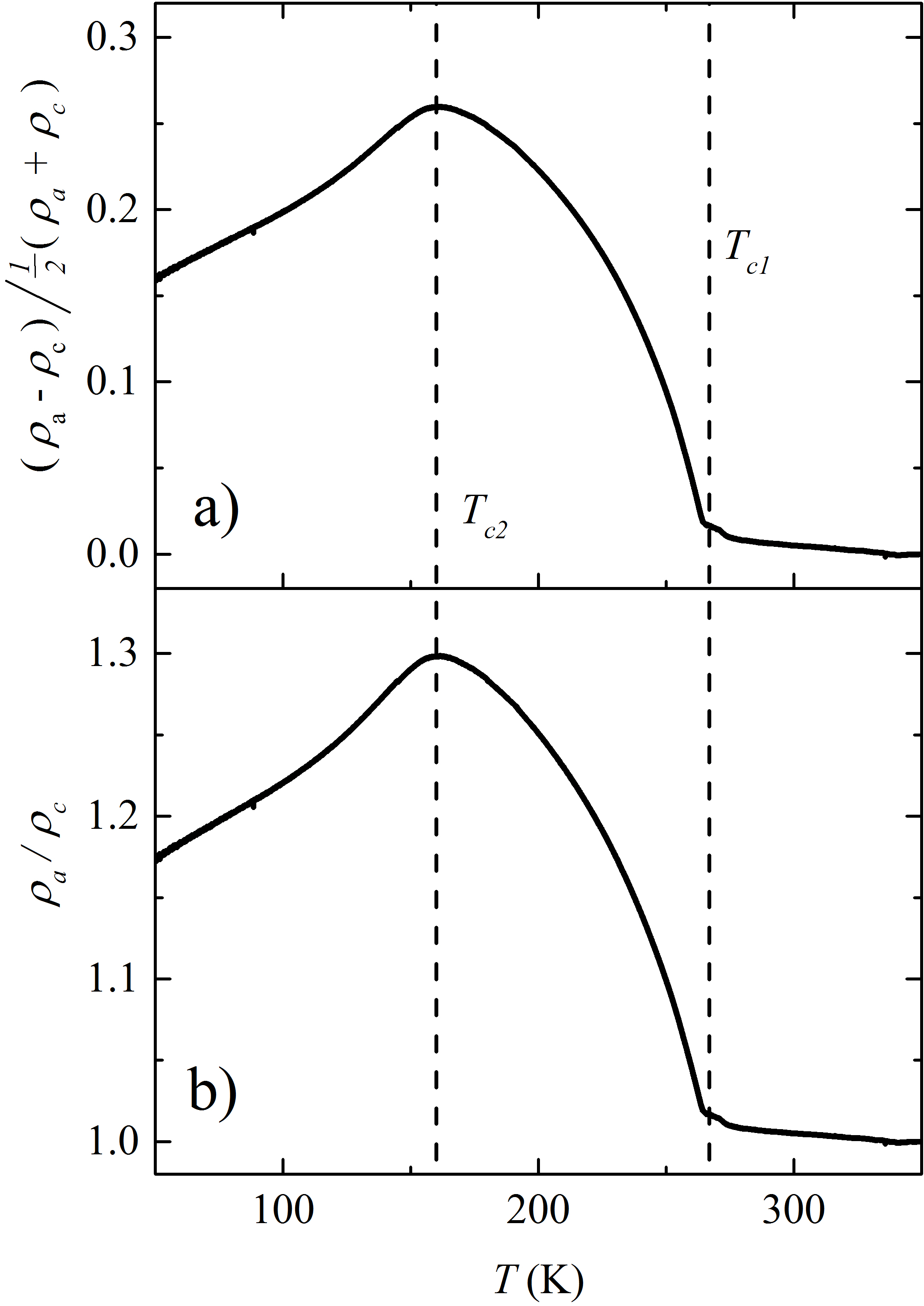}
\caption{a) The figure of merit for resistivity ansiotropy defined as $\frac{\rho_{a}-\rho_{c}}{\nicefrac{1}{2}(\rho_{a}+\rho_{c})}$ shows clear features that coincide with known transition temperatures $T_{c1}$ and $T_{c2}$ (shown as dashed lines). An alternative figure of merit, $\frac{\rho_{a}}{\rho_{c}}$, also highlights very clear features at $T_{c1}$ and $T_{c2}$, as shown in b). For small deviations from the average value these two figures of merit should have the same temperature dependence, consistent with the data.}
\label{fig:rhoa_over_rhoc}
\end{figure}

Two commonly used figures of merit for resistivity anisotropy, $\frac{\rho_{a}-\rho_{c}}{\nicefrac{1}{2}(\rho_{a}+\rho_{c})}$ and $\frac{\rho_{a}}{\rho_{c}}$, are shown in Figures \ref{fig:rhoa_over_rhoc}a and \ref{fig:rhoa_over_rhoc}b respectively; both CDW transitions are easily identified in either plot with the inferred transition temperatures consistent with published values\cite{ru2008chempress,moore2010ErTe3}. For small deviations from the average both of these figures of merit should observe the same temperature dependence, which is consistent with the data. It should be noted that resistivity has a non-trivial relationship to the CDW order parameter, and so $T_{c1}$ indicated in Figures \ref{fig:analysed_data} and \ref{fig:rhoa_over_rhoc} for comparison is instead taken from x-ray measurements of the associated integrated superlattice peak intensity, with the square root of this value being an appropriate order parameter\cite{ru2008chempress}. An equivalent data set of sufficient quality is not available for $T_{c2}$, and so this was derived from ARPES measurements of the energy gap on the Fermi-surface which should be a good proxy for the order parameter\cite{moore2010ErTe3}.

\section{\label{sec:discussion}Discussion}
The transverse method presented here has a number of advantages over both the two-bar method and the Montgomery method when considering the errors already discussed in section \ref{sec:methods}, predominantly because the technique provides a direct measurement of resistivity anisotropy $(\rho_{ii}-\rho_{jj})$ that does not admix with the average resistivity. The present data shows clearly how measuring $\rho'_{x'x'} = \frac{\rho_{ii}+\rho_{jj}}{2}$ and $\rho'_{x'y'}=\frac{\rho_{ii}-\rho_{jj}}{2}$ (rather than $\rho_{ii}$ and $\rho_{jj}$ separately) is a useful shift in philosophy that allows greater resolution in both relative and absolute values of $(\rho_{ii}-\rho_{jj})$ whilst still yielding good values of $\rho_{ii}$ and $\rho_{jj}$ individually. The key sources of error with this technique are transverse contact alignment and angular alignment errors in $\rho'_{x'x'}$. The latter is not an issue if $(\rho_{ii}-\rho_{jj})$ is the relevant quantity to be found (because angular misalignment does not admix $(\rho_{ii}+\rho_{jj})$), and is minimised in absolute terms if $\rho_{ii}\approx\rho_{jj}$. The contribution of the former is robustly corrected for samples that are known to be isotropic in some accessible regime, such as samples undergoing tetragonal to orthorhombic distortions, but requires some caveats in systems which are anisotropic throughout the range of measurement if accurate absolute values are to be obtained. Since ErTe$_3$ is essentially isotropic above $T_{c1}$ this effect can be eliminated or at least minimised as described above, but this is not generally true for orthorhombic materials. Microlithographic techniques could be employed to minimise contact offset and angular misalignment in such materials by providing extremely small and well aligned contacts. In general these errors are likely to be less significant than those found in conventional techniques, particularly for small samples. We therefore emphasise that the technique is uniquely sensitive to resistivity anisotropy in comparison to conventional methods.

Finally, we highlight the critical importance of having a single-domain sample for accurate measurements of the resistivity anisotropy. The ErTe$_3$ sample measured here grew as a single-domain, as do many other orthorhombic materials, but this is often not the case. Furthermore, resistivity anisotropy is often useful in systems that undergo a C$_4$ to C$_2$ rotational symmetry breaking transition, which necessarily forms domains in the absence of an external de-twinning field. The presence of domains can create a pseudo-symmetry when averaged over macroscopic length scales that masks the intrinsic anisotropies of the crystal structure, thus leading to an erroneous underestimation, or even elimination, of the resistivity anisotropy. It can be possible to de-twin samples in situ using, for example, applied magnetic fields\cite{ruff2012susceptibility,zapf2014persistent,ando2003anisotropic,chu2010plane} or strain\cite{chu2010anisotropy,man2015dai}. Ideally this de-twinning field can then be removed below the transition temperature to obtain a single domain in the absence of an applied field\cite{zapf2014persistent}, but the sample may simply re-twin depending on the nature of the ordered phase, particularly close to the transition temperature\cite{man2015dai}. Measurements performed with the presence of a de-twinning field are not necessarily invalid or even inaccurate provided that the effect of the detwinning field is correctly treated. Taking the example of strain, the geometric distortion of the sample by a de-twinning field has a small effect on the resistivity that should be smooth and slowly varying with temperature when compared to changes in the Fermi surface and\,/\,or scattering with the onset of an anisotropic order parameter or fluctuations. The total response of the resistivity to a given strain is quantified by a material's elastoresistivity tensor, with the elastoresistance of metals generally found to be small at temperatures far from a phase transition\cite{riggs2015URS,chu2012nematic}. The nature of the elastoresistance close to a phase transition depends on the nature of the transition and varies widely between materials. Similarly, for de-twinning magnetic fields the magnetoresistance and Hall effect should be accounted for.

\section{\label{sec:conclusions}Conclusions}
A novel method for measuring resistivity anisotropy in a single sample utilising transverse resistivity in a rotated experimental frame has been presented and contrasted with conventional methods. It is shown through error propagation that the transverse method is far less susceptible to admixing effects between the anisotropic and isotropic components of the resistivity than conventional methods and thus presents a more accurate measure of the resistivity anisotropy (provided that the transverse contact misalignment is accounted for or minimised as described). The technique has been successfully applied to ErTe$_3$, clearly identifying the two CDW transitions from changes in the resistive anisotropy and producing absolute values for $\rho_a$ and $\rho_c$ that are consistent with those already published\cite{sinchenko2014montgomery, ru2008chempress}. The direct measurement of $(\rho_{a}-\rho_{c})$ via the transverse voltage contacts is shown to be very sensitive to changes in anisotropy. When compared to the two-bar method, the other most significant advantage is that the measurement elimintates errors due to inequivalency of the two samples and their measurements that effectively lead to admixing of the average resistivity into the determined resistivity anisotropy. In comparison to the Montgomery method, the key advantage is that the signal to noise is generally over an order of magnitude improved, with admixing effects also reduced in this case. The critical importance of measuring a single-domain sample is also highlighted and discussed. To conclude, in many cases the transverse method should be a substantial improvement on existing methods for measuring resistivity anisotropy in both sensitivity and absolute accuracy.

\section*{Acknowledgments}
The authors would like to thank M. C. Shapiro, S. Aeschlimann, H. J. Silverstein  and A. T. Hristov for enlightening and productive discussions. This work was supported by the DOE, Office of Basic Energy Sciences, under Contract No. DEAC02-76SF00515. P.W. was partially supported by the Gordon and Betty Moore Foundation EPiQS Initiative through grant GBMF4414.

\section*{\label{sec:alt_derivation}Appendix A: Alternative explanation of the origin of the transverse electric field}
In the transverse technique as applied in the main text to ErTe$_3$ the current is oriented along the (1 0 1) direction, thus the current density $\vec{J}$ vector can be defined as the sum of two vectors aligned to the crystallographic axes, $\vec{J}=\frac{1}{\sqrt{2}}|\vec{J}|(\hat{a}+\hat{c})$. For currents parallel to the crystallographic axes the conventional resistivity tensor as defined in Equation \ref{eqn:resistivity_tensor} is appropriate, thus the resultant electric field vector $|\vec{E}|$ becomes,

\begin{equation}
\vec{E}=\frac{1}{\sqrt{2}}|\vec{J}|(\rho_a \hat{a} + \rho_c \hat{c}).
\end{equation}
By inspection, $\vec{J} \parallel \vec{E}$ only if $\rho_a = \rho_c$, therefore if the resistivity anisotropy $(\rho_a - \rho_c)$ is non-zero there must be a component of the electric field perpendicular to the applied current. The electric field parallel to the current is found by,

\begin{equation}
\frac{\vec{J}\cdot\vec{E}}{|\vec{J}|}=\frac{1}{2}|\vec{J}|(\rho_a + \rho_c),
\end{equation}

and the electric field perpendicular to the current by, 

\begin{equation}
\frac{|\vec{J}\times\vec{E}|}{|\vec{J}|}=\frac{1}{2}|\vec{J}|(\rho_a - \rho_c),
\end{equation}

where $(\rho_{a}-\rho_{c})$ is the resistivity anisotropy as described in the main text. This approach is also easily generalised for other planes of measurement.

\bibliography{main.bbl} 

\begin{thebibliography}{44}%
\makeatletter
\providecommand \@ifxundefined [1]{%
 \@ifx{#1\undefined}
}%
\providecommand \@ifnum [1]{%
 \ifnum #1\expandafter \@firstoftwo
 \else \expandafter \@secondoftwo
 \fi
}%
\providecommand \@ifx [1]{%
 \ifx #1\expandafter \@firstoftwo
 \else \expandafter \@secondoftwo
 \fi
}%
\providecommand \natexlab [1]{#1}%
\providecommand \enquote  [1]{``#1''}%
\providecommand \bibnamefont  [1]{#1}%
\providecommand \bibfnamefont [1]{#1}%
\providecommand \citenamefont [1]{#1}%
\providecommand \href@noop [0]{\@secondoftwo}%
\providecommand \href [0]{\begingroup \@sanitize@url \@href}%
\providecommand \@href[1]{\@@startlink{#1}\@@href}%
\providecommand \@@href[1]{\endgroup#1\@@endlink}%
\providecommand \@sanitize@url [0]{\catcode `\\12\catcode `\$12\catcode
  `\&12\catcode `\#12\catcode `\^12\catcode `\_12\catcode `\%12\relax}%
\providecommand \@@startlink[1]{}%
\providecommand \@@endlink[0]{}%
\providecommand \url  [0]{\begingroup\@sanitize@url \@url }%
\providecommand \@url [1]{\endgroup\@href {#1}{\urlprefix }}%
\providecommand \urlprefix  [0]{URL }%
\providecommand \Eprint [0]{\href }%
\providecommand \doibase [0]{http://dx.doi.org/}%
\providecommand \selectlanguage [0]{\@gobble}%
\providecommand \bibinfo  [0]{\@secondoftwo}%
\providecommand \bibfield  [0]{\@secondoftwo}%
\providecommand \translation [1]{[#1]}%
\providecommand \BibitemOpen [0]{}%
\providecommand \bibitemStop [0]{}%
\providecommand \bibitemNoStop [0]{.\EOS\space}%
\providecommand \EOS [0]{\spacefactor3000\relax}%
\providecommand \BibitemShut  [1]{\csname bibitem#1\endcsname}%
\let\auto@bib@innerbib\@empty
\bibitem [{\citenamefont {Fradkin}\ \emph {et~al.}(2010)\citenamefont
  {Fradkin}, \citenamefont {Kivelson}, \citenamefont {Lawler}, \citenamefont
  {Eisenstein},\ and\ \citenamefont {Mackenzie}}]{fradkin2009nematic}%
  \BibitemOpen
  \bibfield  {author} {\bibinfo {author} {\bibfnamefont {E.}~\bibnamefont
  {Fradkin}}, \bibinfo {author} {\bibfnamefont {S.~A.}\ \bibnamefont
  {Kivelson}}, \bibinfo {author} {\bibfnamefont {M.~J.}\ \bibnamefont
  {Lawler}}, \bibinfo {author} {\bibfnamefont {J.~P.}\ \bibnamefont
  {Eisenstein}}, \ and\ \bibinfo {author} {\bibfnamefont {A.~P.}\ \bibnamefont
  {Mackenzie}},\ }\href@noop {} {\bibfield  {journal} {\bibinfo  {journal}
  {Annual Review of Condensed Matter Physics}\ }\textbf {\bibinfo {volume}
  {1}},\ \bibinfo {pages} {153} (\bibinfo {year} {2010})}\BibitemShut {NoStop}%
\bibitem [{\citenamefont {Kivelson}, \citenamefont {Fradkin},\ and\
  \citenamefont {Emery}(1998)}]{kivelson1998electronic}%
  \BibitemOpen
  \bibfield  {author} {\bibinfo {author} {\bibfnamefont {S.~A.}\ \bibnamefont
  {Kivelson}}, \bibinfo {author} {\bibfnamefont {E.}~\bibnamefont {Fradkin}}, \
  and\ \bibinfo {author} {\bibfnamefont {V.~J.}\ \bibnamefont {Emery}},\
  }\href@noop {} {\bibfield  {journal} {\bibinfo  {journal} {Nature}\ }\textbf
  {\bibinfo {volume} {393}},\ \bibinfo {pages} {550} (\bibinfo {year}
  {1998})}\BibitemShut {NoStop}%
\bibitem [{pro()}]{proportionalitycomment}%
  \BibitemOpen
  \href@noop {} {}\bibinfo {note} {This proportionality is only strictly true
  in the asymptotic limit, but has been shown to be valid even to large values
  of the order parameter (see following references).}\BibitemShut {Stop}%
\bibitem [{\citenamefont {Carlson}\ \emph {et~al.}(2006)\citenamefont
  {Carlson}, \citenamefont {Dahmen}, \citenamefont {Fradkin},\ and\
  \citenamefont {Kivelson}}]{carlson2006hysteresis}%
  \BibitemOpen
  \bibfield  {author} {\bibinfo {author} {\bibfnamefont {E.~W.}\ \bibnamefont
  {Carlson}}, \bibinfo {author} {\bibfnamefont {K.~A.}\ \bibnamefont {Dahmen}},
  \bibinfo {author} {\bibfnamefont {E.}~\bibnamefont {Fradkin}}, \ and\
  \bibinfo {author} {\bibfnamefont {S.~A.}\ \bibnamefont {Kivelson}},\
  }\href@noop {} {\bibfield  {journal} {\bibinfo  {journal} {Physical review
  letters}\ }\textbf {\bibinfo {volume} {96}},\ \bibinfo {pages} {097003}
  (\bibinfo {year} {2006})}\BibitemShut {NoStop}%
\bibitem [{\citenamefont {Fernandes}, \citenamefont {Abrahams},\ and\
  \citenamefont {Schmalian}(2011)}]{fernandes2011anisotropicres}%
  \BibitemOpen
  \bibfield  {author} {\bibinfo {author} {\bibfnamefont {R.~M.}\ \bibnamefont
  {Fernandes}}, \bibinfo {author} {\bibfnamefont {E.}~\bibnamefont {Abrahams}},
  \ and\ \bibinfo {author} {\bibfnamefont {J.}~\bibnamefont {Schmalian}},\
  }\href@noop {} {\bibfield  {journal} {\bibinfo  {journal} {Physical Review
  Letters}\ }\textbf {\bibinfo {volume} {107}},\ \bibinfo {pages} {217002}
  (\bibinfo {year} {2011})}\BibitemShut {NoStop}%
\bibitem [{\citenamefont {Blomberg}\ \emph {et~al.}(2012)\citenamefont
  {Blomberg}, \citenamefont {Kreyssig}, \citenamefont {Tanatar}, \citenamefont
  {Fernandes}, \citenamefont {Kim}, \citenamefont {Thaler}, \citenamefont
  {Schmalian}, \citenamefont {Bud'ko}, \citenamefont {Canfield}, \citenamefont
  {Goldman},\ and\ \citenamefont {Prozorov}}]{blomberg2012effect}%
  \BibitemOpen
  \bibfield  {author} {\bibinfo {author} {\bibfnamefont {E.~C.}\ \bibnamefont
  {Blomberg}}, \bibinfo {author} {\bibfnamefont {A.}~\bibnamefont {Kreyssig}},
  \bibinfo {author} {\bibfnamefont {M.~A.}\ \bibnamefont {Tanatar}}, \bibinfo
  {author} {\bibfnamefont {R.~M.}\ \bibnamefont {Fernandes}}, \bibinfo {author}
  {\bibfnamefont {M.~G.}\ \bibnamefont {Kim}}, \bibinfo {author} {\bibfnamefont
  {A.}~\bibnamefont {Thaler}}, \bibinfo {author} {\bibfnamefont
  {J.}~\bibnamefont {Schmalian}}, \bibinfo {author} {\bibfnamefont {S.~L.}\
  \bibnamefont {Bud'ko}}, \bibinfo {author} {\bibfnamefont {P.~C.}\
  \bibnamefont {Canfield}}, \bibinfo {author} {\bibfnamefont {A.~I.}\
  \bibnamefont {Goldman}}, \ and\ \bibinfo {author} {\bibfnamefont
  {R.}~\bibnamefont {Prozorov}},\ }\href@noop {} {\bibfield  {journal}
  {\bibinfo  {journal} {Physical Review B}\ }\textbf {\bibinfo {volume} {85}},\
  \bibinfo {pages} {144509} (\bibinfo {year} {2012})}\BibitemShut {NoStop}%
\bibitem [{\citenamefont {Fisher}, \citenamefont {Degiorgi},\ and\
  \citenamefont {Shen}(2011)}]{fisher2011plane}%
  \BibitemOpen
  \bibfield  {author} {\bibinfo {author} {\bibfnamefont {I.~R.}\ \bibnamefont
  {Fisher}}, \bibinfo {author} {\bibfnamefont {L.}~\bibnamefont {Degiorgi}}, \
  and\ \bibinfo {author} {\bibfnamefont {Z.~X.}\ \bibnamefont {Shen}},\
  }\href@noop {} {\bibfield  {journal} {\bibinfo  {journal} {Reports on
  progress in Physics}\ }\textbf {\bibinfo {volume} {74}},\ \bibinfo {pages}
  {124506} (\bibinfo {year} {2011})}\BibitemShut {NoStop}%
\bibitem [{\citenamefont {Chu}\ \emph {et~al.}(2010{\natexlab{a}})\citenamefont
  {Chu}, \citenamefont {Analytis}, \citenamefont {De~Greve}, \citenamefont
  {McMahon}, \citenamefont {Islam}, \citenamefont {Yamamoto},\ and\
  \citenamefont {Fisher}}]{chu2010anisotropy}%
  \BibitemOpen
  \bibfield  {author} {\bibinfo {author} {\bibfnamefont {J.-H.}\ \bibnamefont
  {Chu}}, \bibinfo {author} {\bibfnamefont {J.~G.}\ \bibnamefont {Analytis}},
  \bibinfo {author} {\bibfnamefont {K.}~\bibnamefont {De~Greve}}, \bibinfo
  {author} {\bibfnamefont {P.~L.}\ \bibnamefont {McMahon}}, \bibinfo {author}
  {\bibfnamefont {Z.}~\bibnamefont {Islam}}, \bibinfo {author} {\bibfnamefont
  {Y.}~\bibnamefont {Yamamoto}}, \ and\ \bibinfo {author} {\bibfnamefont
  {I.~R.}\ \bibnamefont {Fisher}},\ }\href@noop {} {\bibfield  {journal}
  {\bibinfo  {journal} {Science}\ }\textbf {\bibinfo {volume} {329}},\ \bibinfo
  {pages} {824} (\bibinfo {year} {2010}{\natexlab{a}})}\BibitemShut {NoStop}%
\bibitem [{\citenamefont {Man}\ \emph {et~al.}(2015)\citenamefont {Man},
  \citenamefont {Lu}, \citenamefont {Chen}, \citenamefont {Zhang},
  \citenamefont {Zhang}, \citenamefont {Luo}, \citenamefont {Kulda},
  \citenamefont {Ivanov}, \citenamefont {Keller}, \citenamefont {Morosan},
  \citenamefont {Si},\ and\ \citenamefont {Dai}}]{man2015dai}%
  \BibitemOpen
  \bibfield  {author} {\bibinfo {author} {\bibfnamefont {H.}~\bibnamefont
  {Man}}, \bibinfo {author} {\bibfnamefont {X.}~\bibnamefont {Lu}}, \bibinfo
  {author} {\bibfnamefont {J.~S.}\ \bibnamefont {Chen}}, \bibinfo {author}
  {\bibfnamefont {R.}~\bibnamefont {Zhang}}, \bibinfo {author} {\bibfnamefont
  {W.}~\bibnamefont {Zhang}}, \bibinfo {author} {\bibfnamefont
  {H.}~\bibnamefont {Luo}}, \bibinfo {author} {\bibfnamefont {J.}~\bibnamefont
  {Kulda}}, \bibinfo {author} {\bibfnamefont {A.}~\bibnamefont {Ivanov}},
  \bibinfo {author} {\bibfnamefont {T.}~\bibnamefont {Keller}}, \bibinfo
  {author} {\bibfnamefont {E.}~\bibnamefont {Morosan}}, \bibinfo {author}
  {\bibfnamefont {Q.}~\bibnamefont {Si}}, \ and\ \bibinfo {author}
  {\bibfnamefont {P.}~\bibnamefont {Dai}},\ }\href@noop {} {\bibfield
  {journal} {\bibinfo  {journal} {Physical Review B}\ }\textbf {\bibinfo
  {volume} {92}},\ \bibinfo {pages} {134521} (\bibinfo {year}
  {2015})}\BibitemShut {NoStop}%
\bibitem [{\citenamefont {Tanatar}\ \emph {et~al.}(2016)\citenamefont
  {Tanatar}, \citenamefont {B{\"o}hmer}, \citenamefont {Timmons}, \citenamefont
  {Sch{\"u}tt}, \citenamefont {Drachuck}, \citenamefont {Taufour},
  \citenamefont {Bud'ko}, \citenamefont {Canfield}, \citenamefont {Fernandes},\
  and\ \citenamefont {Prozorov}}]{tanatar2015fese}%
  \BibitemOpen
  \bibfield  {author} {\bibinfo {author} {\bibfnamefont {M.~A.}\ \bibnamefont
  {Tanatar}}, \bibinfo {author} {\bibfnamefont {A.~E.}\ \bibnamefont
  {B{\"o}hmer}}, \bibinfo {author} {\bibfnamefont {E.~I.}\ \bibnamefont
  {Timmons}}, \bibinfo {author} {\bibfnamefont {M.}~\bibnamefont {Sch{\"u}tt}},
  \bibinfo {author} {\bibfnamefont {G.}~\bibnamefont {Drachuck}}, \bibinfo
  {author} {\bibfnamefont {V.}~\bibnamefont {Taufour}}, \bibinfo {author}
  {\bibfnamefont {S.~L.}\ \bibnamefont {Bud'ko}}, \bibinfo {author}
  {\bibfnamefont {P.~C.}\ \bibnamefont {Canfield}}, \bibinfo {author}
  {\bibfnamefont {R.~M.}\ \bibnamefont {Fernandes}}, \ and\ \bibinfo {author}
  {\bibfnamefont {R.}~\bibnamefont {Prozorov}},\ }\href@noop {} {\bibfield
  {journal} {\bibinfo  {journal} {Physical Review Letters}\ }\textbf {\bibinfo
  {volume} {117}},\ \bibinfo {pages} {127001} (\bibinfo {year}
  {2016})}\BibitemShut {NoStop}%
\bibitem [{\citenamefont {Tanatar}\ \emph {et~al.}(2010)\citenamefont
  {Tanatar}, \citenamefont {Blomberg}, \citenamefont {Kreyssig}, \citenamefont
  {Kim}, \citenamefont {Ni}, \citenamefont {Thaler}, \citenamefont {Bud’ko},
  \citenamefont {Canfield}, \citenamefont {Goldman}, \citenamefont {Mazin},\
  and\ \citenamefont {Prozorov}}]{tanatar2010uniaxial}%
  \BibitemOpen
  \bibfield  {author} {\bibinfo {author} {\bibfnamefont {M.~A.}\ \bibnamefont
  {Tanatar}}, \bibinfo {author} {\bibfnamefont {E.~C.}\ \bibnamefont
  {Blomberg}}, \bibinfo {author} {\bibfnamefont {A.}~\bibnamefont {Kreyssig}},
  \bibinfo {author} {\bibfnamefont {M.~G.}\ \bibnamefont {Kim}}, \bibinfo
  {author} {\bibfnamefont {N.}~\bibnamefont {Ni}}, \bibinfo {author}
  {\bibfnamefont {A.}~\bibnamefont {Thaler}}, \bibinfo {author} {\bibfnamefont
  {S.~L.}\ \bibnamefont {Bud’ko}}, \bibinfo {author} {\bibfnamefont {P.~C.}\
  \bibnamefont {Canfield}}, \bibinfo {author} {\bibfnamefont {A.~I.}\
  \bibnamefont {Goldman}}, \bibinfo {author} {\bibfnamefont {I.~I.}\
  \bibnamefont {Mazin}}, \ and\ \bibinfo {author} {\bibfnamefont
  {R.}~\bibnamefont {Prozorov}},\ }\href@noop {} {\bibfield  {journal}
  {\bibinfo  {journal} {Physical Review B}\ }\textbf {\bibinfo {volume} {81}},\
  \bibinfo {pages} {184508} (\bibinfo {year} {2010})}\BibitemShut {NoStop}%
\bibitem [{\citenamefont {Blomberg}\ \emph {et~al.}(2011)\citenamefont
  {Blomberg}, \citenamefont {Tanatar}, \citenamefont {Kreyssig}, \citenamefont
  {Ni}, \citenamefont {Thaler}, \citenamefont {Hu}, \citenamefont {Bud'ko},
  \citenamefont {Canfield}, \citenamefont {Goldman},\ and\ \citenamefont
  {Prozorov}}]{blomberg2011srfeas}%
  \BibitemOpen
  \bibfield  {author} {\bibinfo {author} {\bibfnamefont {E.~C.}\ \bibnamefont
  {Blomberg}}, \bibinfo {author} {\bibfnamefont {M.~A.}\ \bibnamefont
  {Tanatar}}, \bibinfo {author} {\bibfnamefont {A.}~\bibnamefont {Kreyssig}},
  \bibinfo {author} {\bibfnamefont {N.}~\bibnamefont {Ni}}, \bibinfo {author}
  {\bibfnamefont {A.}~\bibnamefont {Thaler}}, \bibinfo {author} {\bibfnamefont
  {R.}~\bibnamefont {Hu}}, \bibinfo {author} {\bibfnamefont {S.~L.}\
  \bibnamefont {Bud'ko}}, \bibinfo {author} {\bibfnamefont {P.~C.}\
  \bibnamefont {Canfield}}, \bibinfo {author} {\bibfnamefont {A.~I.}\
  \bibnamefont {Goldman}}, \ and\ \bibinfo {author} {\bibfnamefont
  {R.}~\bibnamefont {Prozorov}},\ }\href {\doibase 10.1103/PhysRevB.83.134505}
  {\bibfield  {journal} {\bibinfo  {journal} {Phys. Rev. B}\ }\textbf {\bibinfo
  {volume} {83}},\ \bibinfo {pages} {134505} (\bibinfo {year}
  {2011})}\BibitemShut {NoStop}%
\bibitem [{\citenamefont {Kuo}\ \emph {et~al.}(2011)\citenamefont {Kuo},
  \citenamefont {Chu}, \citenamefont {Riggs}, \citenamefont {Yu}, \citenamefont
  {McMahon}, \citenamefont {De~Greve}, \citenamefont {Yamamoto}, \citenamefont
  {Analytis},\ and\ \citenamefont {Fisher}}]{kuo2011possible}%
  \BibitemOpen
  \bibfield  {author} {\bibinfo {author} {\bibfnamefont {H.-H.}\ \bibnamefont
  {Kuo}}, \bibinfo {author} {\bibfnamefont {J.-H.}\ \bibnamefont {Chu}},
  \bibinfo {author} {\bibfnamefont {S.~C.}\ \bibnamefont {Riggs}}, \bibinfo
  {author} {\bibfnamefont {L.}~\bibnamefont {Yu}}, \bibinfo {author}
  {\bibfnamefont {P.~L.}\ \bibnamefont {McMahon}}, \bibinfo {author}
  {\bibfnamefont {K.}~\bibnamefont {De~Greve}}, \bibinfo {author}
  {\bibfnamefont {Y.}~\bibnamefont {Yamamoto}}, \bibinfo {author}
  {\bibfnamefont {J.~G.}\ \bibnamefont {Analytis}}, \ and\ \bibinfo {author}
  {\bibfnamefont {I.~R.}\ \bibnamefont {Fisher}},\ }\href@noop {} {\bibfield
  {journal} {\bibinfo  {journal} {Physical Review B}\ }\textbf {\bibinfo
  {volume} {84}},\ \bibinfo {pages} {054540} (\bibinfo {year}
  {2011})}\BibitemShut {NoStop}%
\bibitem [{\citenamefont {Liang}\ \emph {et~al.}(2011)\citenamefont {Liang},
  \citenamefont {Nakajima}, \citenamefont {Kihou}, \citenamefont {Tomioka},
  \citenamefont {Ito}, \citenamefont {Lee}, \citenamefont {Kito}, \citenamefont
  {Iyo}, \citenamefont {Eisaki}, \citenamefont {Kakeshita},\ and\ \citenamefont
  {Uchida}}]{liang2011effects}%
  \BibitemOpen
  \bibfield  {author} {\bibinfo {author} {\bibfnamefont {T.}~\bibnamefont
  {Liang}}, \bibinfo {author} {\bibfnamefont {M.}~\bibnamefont {Nakajima}},
  \bibinfo {author} {\bibfnamefont {K.}~\bibnamefont {Kihou}}, \bibinfo
  {author} {\bibfnamefont {Y.}~\bibnamefont {Tomioka}}, \bibinfo {author}
  {\bibfnamefont {T.}~\bibnamefont {Ito}}, \bibinfo {author} {\bibfnamefont
  {C.~H.}\ \bibnamefont {Lee}}, \bibinfo {author} {\bibfnamefont
  {H.}~\bibnamefont {Kito}}, \bibinfo {author} {\bibfnamefont {A.}~\bibnamefont
  {Iyo}}, \bibinfo {author} {\bibfnamefont {H.}~\bibnamefont {Eisaki}},
  \bibinfo {author} {\bibfnamefont {T.}~\bibnamefont {Kakeshita}}, \ and\
  \bibinfo {author} {\bibfnamefont {S.}~\bibnamefont {Uchida}},\ }\href@noop {}
  {\bibfield  {journal} {\bibinfo  {journal} {Journal of Physics and Chemistry
  of Solids}\ }\textbf {\bibinfo {volume} {72}},\ \bibinfo {pages} {418}
  (\bibinfo {year} {2011})}\BibitemShut {NoStop}%
\bibitem [{\citenamefont {Ying}\ \emph {et~al.}(2011)\citenamefont {Ying},
  \citenamefont {Wang}, \citenamefont {Wu}, \citenamefont {Xiang},
  \citenamefont {Liu}, \citenamefont {Yan}, \citenamefont {Wang}, \citenamefont
  {Zhang}, \citenamefont {Ye}, \citenamefont {Cheng}, \citenamefont {Hu},\ and\
  \citenamefont {Chen}}]{ying2011measurements}%
  \BibitemOpen
  \bibfield  {author} {\bibinfo {author} {\bibfnamefont {J.~J.}\ \bibnamefont
  {Ying}}, \bibinfo {author} {\bibfnamefont {X.~F.}\ \bibnamefont {Wang}},
  \bibinfo {author} {\bibfnamefont {T.}~\bibnamefont {Wu}}, \bibinfo {author}
  {\bibfnamefont {Z.~J.}\ \bibnamefont {Xiang}}, \bibinfo {author}
  {\bibfnamefont {R.~H.}\ \bibnamefont {Liu}}, \bibinfo {author} {\bibfnamefont
  {Y.~J.}\ \bibnamefont {Yan}}, \bibinfo {author} {\bibfnamefont {A.~F.}\
  \bibnamefont {Wang}}, \bibinfo {author} {\bibfnamefont {M.}~\bibnamefont
  {Zhang}}, \bibinfo {author} {\bibfnamefont {G.~J.}\ \bibnamefont {Ye}},
  \bibinfo {author} {\bibfnamefont {P.}~\bibnamefont {Cheng}}, \bibinfo
  {author} {\bibfnamefont {J.~P.}\ \bibnamefont {Hu}}, \ and\ \bibinfo {author}
  {\bibfnamefont {X.~H.}\ \bibnamefont {Chen}},\ }\href@noop {} {\bibfield
  {journal} {\bibinfo  {journal} {Physical review letters}\ }\textbf {\bibinfo
  {volume} {107}},\ \bibinfo {pages} {067001} (\bibinfo {year}
  {2011})}\BibitemShut {NoStop}%
\bibitem [{\citenamefont {Jiang}\ \emph {et~al.}(2012)\citenamefont {Jiang},
  \citenamefont {He}, \citenamefont {Zhang}, \citenamefont {Xu}, \citenamefont
  {Ge}, \citenamefont {Ye}, \citenamefont {Chen}, \citenamefont {Xie},\ and\
  \citenamefont {Feng}}]{Jiang2012anisotropy}%
  \BibitemOpen
  \bibfield  {author} {\bibinfo {author} {\bibfnamefont {J.}~\bibnamefont
  {Jiang}}, \bibinfo {author} {\bibfnamefont {C.}~\bibnamefont {He}}, \bibinfo
  {author} {\bibfnamefont {Y.}~\bibnamefont {Zhang}}, \bibinfo {author}
  {\bibfnamefont {M.}~\bibnamefont {Xu}}, \bibinfo {author} {\bibfnamefont
  {Q.~Q.}\ \bibnamefont {Ge}}, \bibinfo {author} {\bibfnamefont {Z.~R.}\
  \bibnamefont {Ye}}, \bibinfo {author} {\bibfnamefont {F.}~\bibnamefont
  {Chen}}, \bibinfo {author} {\bibfnamefont {B.~P.}\ \bibnamefont {Xie}}, \
  and\ \bibinfo {author} {\bibfnamefont {D.~L.}\ \bibnamefont {Feng}},\
  }\href@noop {} {\bibfield  {journal} {\bibinfo  {journal} {arXiv:1210.0397}\
  } (\bibinfo {year} {2012})}\BibitemShut {NoStop}%
\bibitem [{\citenamefont {Chu}\ \emph {et~al.}(2012)\citenamefont {Chu},
  \citenamefont {Kuo}, \citenamefont {Analytis},\ and\ \citenamefont
  {Fisher}}]{chu2012nematic}%
  \BibitemOpen
  \bibfield  {author} {\bibinfo {author} {\bibfnamefont {J.-H.}\ \bibnamefont
  {Chu}}, \bibinfo {author} {\bibfnamefont {H.-H.}\ \bibnamefont {Kuo}},
  \bibinfo {author} {\bibfnamefont {J.~G.}\ \bibnamefont {Analytis}}, \ and\
  \bibinfo {author} {\bibfnamefont {I.~R.}\ \bibnamefont {Fisher}},\
  }\href@noop {} {\bibfield  {journal} {\bibinfo  {journal} {Science}\ }\textbf
  {\bibinfo {volume} {337}},\ \bibinfo {pages} {710} (\bibinfo {year}
  {2012})}\BibitemShut {NoStop}%
\bibitem [{\citenamefont {Kuo}\ \emph {et~al.}(2016)\citenamefont {Kuo},
  \citenamefont {Chu}, \citenamefont {Palmstrom}, \citenamefont {Kivelson},\
  and\ \citenamefont {Fisher}}]{kuo2016ubiquitous}%
  \BibitemOpen
  \bibfield  {author} {\bibinfo {author} {\bibfnamefont {H.-H.}\ \bibnamefont
  {Kuo}}, \bibinfo {author} {\bibfnamefont {J.-H.}\ \bibnamefont {Chu}},
  \bibinfo {author} {\bibfnamefont {J.~C.}\ \bibnamefont {Palmstrom}}, \bibinfo
  {author} {\bibfnamefont {S.~A.}\ \bibnamefont {Kivelson}}, \ and\ \bibinfo
  {author} {\bibfnamefont {I.~R.}\ \bibnamefont {Fisher}},\ }\href@noop {}
  {\bibfield  {journal} {\bibinfo  {journal} {Science}\ }\textbf {\bibinfo
  {volume} {352}},\ \bibinfo {pages} {958} (\bibinfo {year}
  {2016})}\BibitemShut {NoStop}%
\bibitem [{\citenamefont {Watson}\ \emph {et~al.}(2015)\citenamefont {Watson},
  \citenamefont {Kim}, \citenamefont {Haghighirad}, \citenamefont {Davies},
  \citenamefont {McCollam}, \citenamefont {Narayanan}, \citenamefont {Blake},
  \citenamefont {Chen}, \citenamefont {Ghannadzadeh}, \citenamefont
  {Schofield}, \citenamefont {Hoesch}, \citenamefont {Meingast}, \citenamefont
  {Wolf},\ and\ \citenamefont {Coldea}}]{watson2015fese}%
  \BibitemOpen
  \bibfield  {author} {\bibinfo {author} {\bibfnamefont {M.~D.}\ \bibnamefont
  {Watson}}, \bibinfo {author} {\bibfnamefont {T.~K.}\ \bibnamefont {Kim}},
  \bibinfo {author} {\bibfnamefont {A.~A.}\ \bibnamefont {Haghighirad}},
  \bibinfo {author} {\bibfnamefont {N.~R.}\ \bibnamefont {Davies}}, \bibinfo
  {author} {\bibfnamefont {A.}~\bibnamefont {McCollam}}, \bibinfo {author}
  {\bibfnamefont {A.}~\bibnamefont {Narayanan}}, \bibinfo {author}
  {\bibfnamefont {S.~F.}\ \bibnamefont {Blake}}, \bibinfo {author}
  {\bibfnamefont {Y.~L.}\ \bibnamefont {Chen}}, \bibinfo {author}
  {\bibfnamefont {S.}~\bibnamefont {Ghannadzadeh}}, \bibinfo {author}
  {\bibfnamefont {A.~J.}\ \bibnamefont {Schofield}}, \bibinfo {author}
  {\bibfnamefont {M.}~\bibnamefont {Hoesch}}, \bibinfo {author} {\bibfnamefont
  {C.}~\bibnamefont {Meingast}}, \bibinfo {author} {\bibfnamefont
  {T.}~\bibnamefont {Wolf}}, \ and\ \bibinfo {author} {\bibfnamefont {A.~I.}\
  \bibnamefont {Coldea}},\ }\href@noop {} {\bibfield  {journal} {\bibinfo
  {journal} {Physical Review B}\ }\textbf {\bibinfo {volume} {91}},\ \bibinfo
  {pages} {155106} (\bibinfo {year} {2015})}\BibitemShut {NoStop}%
\bibitem [{\citenamefont {Kuo}\ \emph {et~al.}(2013)\citenamefont {Kuo},
  \citenamefont {Shapiro}, \citenamefont {Riggs},\ and\ \citenamefont
  {Fisher}}]{kuo2013measurement}%
  \BibitemOpen
  \bibfield  {author} {\bibinfo {author} {\bibfnamefont {H.-H.}\ \bibnamefont
  {Kuo}}, \bibinfo {author} {\bibfnamefont {M.~C.}\ \bibnamefont {Shapiro}},
  \bibinfo {author} {\bibfnamefont {S.~C.}\ \bibnamefont {Riggs}}, \ and\
  \bibinfo {author} {\bibfnamefont {I.~R.}\ \bibnamefont {Fisher}},\
  }\href@noop {} {\bibfield  {journal} {\bibinfo  {journal} {Physical Review
  B}\ }\textbf {\bibinfo {volume} {88}},\ \bibinfo {pages} {085113} (\bibinfo
  {year} {2013})}\BibitemShut {NoStop}%
\bibitem [{\citenamefont {Kuo}\ and\ \citenamefont
  {Fisher}(2014)}]{kuo2014effect}%
  \BibitemOpen
  \bibfield  {author} {\bibinfo {author} {\bibfnamefont {H.-H.}\ \bibnamefont
  {Kuo}}\ and\ \bibinfo {author} {\bibfnamefont {I.~R.}\ \bibnamefont
  {Fisher}},\ }\href@noop {} {\bibfield  {journal} {\bibinfo  {journal}
  {Physical review letters}\ }\textbf {\bibinfo {volume} {112}},\ \bibinfo
  {pages} {227001} (\bibinfo {year} {2014})}\BibitemShut {NoStop}%
\bibitem [{\citenamefont {Shapiro}\ \emph {et~al.}(2016)\citenamefont
  {Shapiro}, \citenamefont {Hristov}, \citenamefont {Palmstrom}, \citenamefont
  {Chu},\ and\ \citenamefont {Fisher}}]{shapiro2016measurement}%
  \BibitemOpen
  \bibfield  {author} {\bibinfo {author} {\bibfnamefont {M.~C.}\ \bibnamefont
  {Shapiro}}, \bibinfo {author} {\bibfnamefont {A.~T.}\ \bibnamefont
  {Hristov}}, \bibinfo {author} {\bibfnamefont {J.~C.}\ \bibnamefont
  {Palmstrom}}, \bibinfo {author} {\bibfnamefont {J.-H.}\ \bibnamefont {Chu}},
  \ and\ \bibinfo {author} {\bibfnamefont {I.~R.}\ \bibnamefont {Fisher}},\
  }\href@noop {} {\bibfield  {journal} {\bibinfo  {journal} {Review of
  Scientific Instruments}\ }\textbf {\bibinfo {volume} {87}},\ \bibinfo {pages}
  {063902} (\bibinfo {year} {2016})}\BibitemShut {NoStop}%
\bibitem [{\citenamefont {Ando}\ \emph {et~al.}(2002)\citenamefont {Ando},
  \citenamefont {Segawa}, \citenamefont {Komiya},\ and\ \citenamefont
  {Lavrov}}]{ando2002YBCO}%
  \BibitemOpen
  \bibfield  {author} {\bibinfo {author} {\bibfnamefont {Y.}~\bibnamefont
  {Ando}}, \bibinfo {author} {\bibfnamefont {K.}~\bibnamefont {Segawa}},
  \bibinfo {author} {\bibfnamefont {S.}~\bibnamefont {Komiya}}, \ and\ \bibinfo
  {author} {\bibfnamefont {A.~N.}\ \bibnamefont {Lavrov}},\ }\href@noop {}
  {\bibfield  {journal} {\bibinfo  {journal} {Physical Review Letters}\
  }\textbf {\bibinfo {volume} {88}},\ \bibinfo {pages} {137005} (\bibinfo
  {year} {2002})}\BibitemShut {NoStop}%
\bibitem [{\citenamefont {Cyr-Choiniere}\ \emph {et~al.}(2015)\citenamefont
  {Cyr-Choiniere}, \citenamefont {Grissonnanche}, \citenamefont {Badoux},
  \citenamefont {Day}, \citenamefont {Bonn}, \citenamefont {Hardy},
  \citenamefont {Liang}, \citenamefont {Doiron-Leyraud},\ and\ \citenamefont
  {Taillefer}}]{cyr2015ybco}%
  \BibitemOpen
  \bibfield  {author} {\bibinfo {author} {\bibfnamefont {O.}~\bibnamefont
  {Cyr-Choiniere}}, \bibinfo {author} {\bibfnamefont {G.}~\bibnamefont
  {Grissonnanche}}, \bibinfo {author} {\bibfnamefont {S.}~\bibnamefont
  {Badoux}}, \bibinfo {author} {\bibfnamefont {J.}~\bibnamefont {Day}},
  \bibinfo {author} {\bibfnamefont {D.}~\bibnamefont {Bonn}}, \bibinfo {author}
  {\bibfnamefont {W.}~\bibnamefont {Hardy}}, \bibinfo {author} {\bibfnamefont
  {R.}~\bibnamefont {Liang}}, \bibinfo {author} {\bibfnamefont
  {N.}~\bibnamefont {Doiron-Leyraud}}, \ and\ \bibinfo {author} {\bibfnamefont
  {L.}~\bibnamefont {Taillefer}},\ }\href@noop {} {\bibfield  {journal}
  {\bibinfo  {journal} {Physical Review B}\ }\textbf {\bibinfo {volume} {92}},\
  \bibinfo {pages} {224502} (\bibinfo {year} {2015})}\BibitemShut {NoStop}%
\bibitem [{\citenamefont {Daou}\ \emph {et~al.}(2010)\citenamefont {Daou},
  \citenamefont {Chang}, \citenamefont {LeBoeuf}, \citenamefont
  {Cyr-Choiniere}, \citenamefont {Lalibert{\'e}}, \citenamefont
  {Doiron-Leyraud}, \citenamefont {Ramshaw}, \citenamefont {Liang},
  \citenamefont {Bonn}, \citenamefont {Hardy},\ and\ \citenamefont
  {Taillefer}}]{daou2010broken}%
  \BibitemOpen
  \bibfield  {author} {\bibinfo {author} {\bibfnamefont {R.}~\bibnamefont
  {Daou}}, \bibinfo {author} {\bibfnamefont {J.}~\bibnamefont {Chang}},
  \bibinfo {author} {\bibfnamefont {D.}~\bibnamefont {LeBoeuf}}, \bibinfo
  {author} {\bibfnamefont {O.}~\bibnamefont {Cyr-Choiniere}}, \bibinfo {author}
  {\bibfnamefont {F.}~\bibnamefont {Lalibert{\'e}}}, \bibinfo {author}
  {\bibfnamefont {N.}~\bibnamefont {Doiron-Leyraud}}, \bibinfo {author}
  {\bibfnamefont {B.~J.}\ \bibnamefont {Ramshaw}}, \bibinfo {author}
  {\bibfnamefont {R.}~\bibnamefont {Liang}}, \bibinfo {author} {\bibfnamefont
  {D.~A.}\ \bibnamefont {Bonn}}, \bibinfo {author} {\bibfnamefont {W.~N.}\
  \bibnamefont {Hardy}}, \ and\ \bibinfo {author} {\bibfnamefont
  {L.}~\bibnamefont {Taillefer}},\ }\href@noop {} {\bibfield  {journal}
  {\bibinfo  {journal} {Nature}\ }\textbf {\bibinfo {volume} {463}},\ \bibinfo
  {pages} {519} (\bibinfo {year} {2010})}\BibitemShut {NoStop}%
\bibitem [{\citenamefont {Chang}\ \emph {et~al.}(2011)\citenamefont {Chang},
  \citenamefont {Doiron-Leyraud}, \citenamefont {Lalibert\'e}, \citenamefont
  {Daou}, \citenamefont {LeBoeuf}, \citenamefont {Ramshaw}, \citenamefont
  {Liang}, \citenamefont {Bonn}, \citenamefont {Hardy}, \citenamefont {Proust},
  \citenamefont {Sheikin}, \citenamefont {Behnia},\ and\ \citenamefont
  {Taillefer}}]{chang2011nernst}%
  \BibitemOpen
  \bibfield  {author} {\bibinfo {author} {\bibfnamefont {J.}~\bibnamefont
  {Chang}}, \bibinfo {author} {\bibfnamefont {N.}~\bibnamefont
  {Doiron-Leyraud}}, \bibinfo {author} {\bibfnamefont {F.}~\bibnamefont
  {Lalibert\'e}}, \bibinfo {author} {\bibfnamefont {R.}~\bibnamefont {Daou}},
  \bibinfo {author} {\bibfnamefont {D.}~\bibnamefont {LeBoeuf}}, \bibinfo
  {author} {\bibfnamefont {B.~J.}\ \bibnamefont {Ramshaw}}, \bibinfo {author}
  {\bibfnamefont {R.}~\bibnamefont {Liang}}, \bibinfo {author} {\bibfnamefont
  {D.~A.}\ \bibnamefont {Bonn}}, \bibinfo {author} {\bibfnamefont {W.~N.}\
  \bibnamefont {Hardy}}, \bibinfo {author} {\bibfnamefont {C.}~\bibnamefont
  {Proust}}, \bibinfo {author} {\bibfnamefont {I.}~\bibnamefont {Sheikin}},
  \bibinfo {author} {\bibfnamefont {K.}~\bibnamefont {Behnia}}, \ and\ \bibinfo
  {author} {\bibfnamefont {L.}~\bibnamefont {Taillefer}},\ }\href {\doibase
  10.1103/PhysRevB.84.014507} {\bibfield  {journal} {\bibinfo  {journal} {Phys.
  Rev. B}\ }\textbf {\bibinfo {volume} {84}},\ \bibinfo {pages} {014507}
  (\bibinfo {year} {2011})}\BibitemShut {NoStop}%
\bibitem [{\citenamefont {Sinchenko}\ \emph {et~al.}(2014)\citenamefont
  {Sinchenko}, \citenamefont {Grigoriev}, \citenamefont {Lejay},\ and\
  \citenamefont {Monceau}}]{sinchenko2014montgomery}%
  \BibitemOpen
  \bibfield  {author} {\bibinfo {author} {\bibfnamefont {A.~A.}\ \bibnamefont
  {Sinchenko}}, \bibinfo {author} {\bibfnamefont {P.~D.}\ \bibnamefont
  {Grigoriev}}, \bibinfo {author} {\bibfnamefont {P.}~\bibnamefont {Lejay}}, \
  and\ \bibinfo {author} {\bibfnamefont {P.}~\bibnamefont {Monceau}},\
  }\href@noop {} {\bibfield  {journal} {\bibinfo  {journal} {Physical Review
  Letters}\ }\textbf {\bibinfo {volume} {112}},\ \bibinfo {pages} {036601}
  (\bibinfo {year} {2014})}\BibitemShut {NoStop}%
\bibitem [{\citenamefont {van~der Pauw}(1958)}]{vanderpauw1958method}%
  \BibitemOpen
  \bibfield  {author} {\bibinfo {author} {\bibfnamefont {L.}~\bibnamefont
  {van~der Pauw}},\ }\href@noop {} {\bibfield  {journal} {\bibinfo  {journal}
  {Philips Research Reports}\ }\textbf {\bibinfo {volume} {13}},\ \bibinfo
  {pages} {1} (\bibinfo {year} {1958})}\BibitemShut {NoStop}%
\bibitem [{\citenamefont {Montgomery}(1971)}]{montgomery1971method}%
  \BibitemOpen
  \bibfield  {author} {\bibinfo {author} {\bibfnamefont {H.~C.}\ \bibnamefont
  {Montgomery}},\ }\href@noop {} {\bibfield  {journal} {\bibinfo  {journal}
  {Journal of Applied Physics}\ }\textbf {\bibinfo {volume} {42}},\ \bibinfo
  {pages} {2971} (\bibinfo {year} {1971})}\BibitemShut {NoStop}%
\bibitem [{\citenamefont {Mercure}\ \emph {et~al.}(2012)\citenamefont
  {Mercure}, \citenamefont {Bangura}, \citenamefont {Xu}, \citenamefont
  {Wakeham}, \citenamefont {Carrington}, \citenamefont {Walmsley},
  \citenamefont {Greenblatt},\ and\ \citenamefont {Hussey}}]{mercure2012lmo}%
  \BibitemOpen
  \bibfield  {author} {\bibinfo {author} {\bibfnamefont {J.-F.}\ \bibnamefont
  {Mercure}}, \bibinfo {author} {\bibfnamefont {A.~F.}\ \bibnamefont
  {Bangura}}, \bibinfo {author} {\bibfnamefont {X.}~\bibnamefont {Xu}},
  \bibinfo {author} {\bibfnamefont {N.}~\bibnamefont {Wakeham}}, \bibinfo
  {author} {\bibfnamefont {A.}~\bibnamefont {Carrington}}, \bibinfo {author}
  {\bibfnamefont {P.}~\bibnamefont {Walmsley}}, \bibinfo {author}
  {\bibfnamefont {M.}~\bibnamefont {Greenblatt}}, \ and\ \bibinfo {author}
  {\bibfnamefont {N.~E.}\ \bibnamefont {Hussey}},\ }\href@noop {} {\bibfield
  {journal} {\bibinfo  {journal} {Physical Review Letters}\ }\textbf {\bibinfo
  {volume} {108}},\ \bibinfo {pages} {187003} (\bibinfo {year}
  {2012})}\BibitemShut {NoStop}%
\bibitem [{\citenamefont {Hussey}\ \emph {et~al.}(2002)\citenamefont {Hussey},
  \citenamefont {McBrien}, \citenamefont {Balicas}, \citenamefont {Brooks},
  \citenamefont {Horii},\ and\ \citenamefont {Ikuta}}]{hussey2002pr124}%
  \BibitemOpen
  \bibfield  {author} {\bibinfo {author} {\bibfnamefont {N.~E.}\ \bibnamefont
  {Hussey}}, \bibinfo {author} {\bibfnamefont {M.~N.}\ \bibnamefont {McBrien}},
  \bibinfo {author} {\bibfnamefont {L.}~\bibnamefont {Balicas}}, \bibinfo
  {author} {\bibfnamefont {J.~S.}\ \bibnamefont {Brooks}}, \bibinfo {author}
  {\bibfnamefont {S.}~\bibnamefont {Horii}}, \ and\ \bibinfo {author}
  {\bibfnamefont {H.}~\bibnamefont {Ikuta}},\ }\href@noop {} {\bibfield
  {journal} {\bibinfo  {journal} {Physical Review Letters}\ }\textbf {\bibinfo
  {volume} {89}},\ \bibinfo {pages} {086601} (\bibinfo {year}
  {2002})}\BibitemShut {NoStop}%
\bibitem [{\citenamefont {Borup}\ \emph {et~al.}(2012)\citenamefont {Borup},
  \citenamefont {Toberer}, \citenamefont {Zoltan}, \citenamefont {Nakatsukasa},
  \citenamefont {Errico}, \citenamefont {Fleurial}, \citenamefont {Iversen},\
  and\ \citenamefont {Snyder}}]{borup2012parallelogram}%
  \BibitemOpen
  \bibfield  {author} {\bibinfo {author} {\bibfnamefont {K.~A.}\ \bibnamefont
  {Borup}}, \bibinfo {author} {\bibfnamefont {E.~S.}\ \bibnamefont {Toberer}},
  \bibinfo {author} {\bibfnamefont {L.~D.}\ \bibnamefont {Zoltan}}, \bibinfo
  {author} {\bibfnamefont {G.}~\bibnamefont {Nakatsukasa}}, \bibinfo {author}
  {\bibfnamefont {M.}~\bibnamefont {Errico}}, \bibinfo {author} {\bibfnamefont
  {J.-P.}\ \bibnamefont {Fleurial}}, \bibinfo {author} {\bibfnamefont {B.~B.}\
  \bibnamefont {Iversen}}, \ and\ \bibinfo {author} {\bibfnamefont {G.~J.}\
  \bibnamefont {Snyder}},\ }\href@noop {} {\bibfield  {journal} {\bibinfo
  {journal} {Review of Scientific Instruments}\ }\textbf {\bibinfo {volume}
  {83}},\ \bibinfo {pages} {123902} (\bibinfo {year} {2012})}\BibitemShut
  {NoStop}%
\bibitem [{\citenamefont {Bierwagen}\ \emph {et~al.}(2004)\citenamefont
  {Bierwagen}, \citenamefont {Pomraenke}, \citenamefont {Eilers},\ and\
  \citenamefont {Masselink}}]{bierwagen2004vdpcalc}%
  \BibitemOpen
  \bibfield  {author} {\bibinfo {author} {\bibfnamefont {O.}~\bibnamefont
  {Bierwagen}}, \bibinfo {author} {\bibfnamefont {R.}~\bibnamefont
  {Pomraenke}}, \bibinfo {author} {\bibfnamefont {S.}~\bibnamefont {Eilers}}, \
  and\ \bibinfo {author} {\bibfnamefont {W.}~\bibnamefont {Masselink}},\
  }\href@noop {} {\bibfield  {journal} {\bibinfo  {journal} {Physical Review
  B}\ }\textbf {\bibinfo {volume} {70}},\ \bibinfo {pages} {165307} (\bibinfo
  {year} {2004})}\BibitemShut {NoStop}%
\bibitem [{nor()}]{normalisationcomment}%
  \BibitemOpen
  \href@noop {} {}\bibinfo {note} {I.e. taking $w$ and $d$ as the width and
  depth of the sample, $\rho'_{x'x'}=R'_{x'x'}\frac{A_{\perp
  x'}}{l_{x'}^{x'}}=\frac{V_{x'}}{I_{x'}}\frac{dw}{l_{x'}^{x'}}$ and
  $\rho'_{x'y'}=R'_{x'y'}d=\frac{V_{y'}}{I_{x'}}d$, provided that the
  transverse contacts are on the edge of the sample such that $l^{y'}_{y'}=w$.
  In this case only a single geometric measurement is required for the
  transverse resistivity, decreasing the associated error contributions. If
  $l^{y'}_{y'}\neq w$ however then
  $\rho'_{x'y'}=\frac{V_{y'}}{I_{x'}}\frac{dw}{l_{y'}^{y'}}$}\BibitemShut
  {NoStop}%
\bibitem [{\citenamefont {Laverock}\ \emph {et~al.}(2005)\citenamefont
  {Laverock}, \citenamefont {Dugdale}, \citenamefont {Major}, \citenamefont
  {Alam}, \citenamefont {Ru}, \citenamefont {Fisher}, \citenamefont {Santi},\
  and\ \citenamefont {Bruno}}]{laverock2005rte3fs}%
  \BibitemOpen
  \bibfield  {author} {\bibinfo {author} {\bibfnamefont {J.}~\bibnamefont
  {Laverock}}, \bibinfo {author} {\bibfnamefont {S.~B.}\ \bibnamefont
  {Dugdale}}, \bibinfo {author} {\bibfnamefont {Z.~S.}\ \bibnamefont {Major}},
  \bibinfo {author} {\bibfnamefont {M.~A.}\ \bibnamefont {Alam}}, \bibinfo
  {author} {\bibfnamefont {N.}~\bibnamefont {Ru}}, \bibinfo {author}
  {\bibfnamefont {I.~R.}\ \bibnamefont {Fisher}}, \bibinfo {author}
  {\bibfnamefont {G.}~\bibnamefont {Santi}}, \ and\ \bibinfo {author}
  {\bibfnamefont {E.}~\bibnamefont {Bruno}},\ }\href@noop {} {\bibfield
  {journal} {\bibinfo  {journal} {Physical Review B}\ }\textbf {\bibinfo
  {volume} {71}},\ \bibinfo {pages} {085114} (\bibinfo {year}
  {2005})}\BibitemShut {NoStop}%
\bibitem [{\citenamefont {Ru}\ \emph {et~al.}(2008)\citenamefont {Ru},
  \citenamefont {Condron}, \citenamefont {Margulis}, \citenamefont {Shin},
  \citenamefont {Laverock}, \citenamefont {Dugdale}, \citenamefont {Toney},\
  and\ \citenamefont {Fisher}}]{ru2008chempress}%
  \BibitemOpen
  \bibfield  {author} {\bibinfo {author} {\bibfnamefont {N.}~\bibnamefont
  {Ru}}, \bibinfo {author} {\bibfnamefont {C.~L.}\ \bibnamefont {Condron}},
  \bibinfo {author} {\bibfnamefont {G.~Y.}\ \bibnamefont {Margulis}}, \bibinfo
  {author} {\bibfnamefont {K.~Y.}\ \bibnamefont {Shin}}, \bibinfo {author}
  {\bibfnamefont {J.}~\bibnamefont {Laverock}}, \bibinfo {author}
  {\bibfnamefont {S.~B.}\ \bibnamefont {Dugdale}}, \bibinfo {author}
  {\bibfnamefont {M.~F.}\ \bibnamefont {Toney}}, \ and\ \bibinfo {author}
  {\bibfnamefont {I.~R.}\ \bibnamefont {Fisher}},\ }\href@noop {} {\bibfield
  {journal} {\bibinfo  {journal} {Physical Review B}\ }\textbf {\bibinfo
  {volume} {77}},\ \bibinfo {pages} {035114} (\bibinfo {year}
  {2008})}\BibitemShut {NoStop}%
\bibitem [{\citenamefont {Maschek}\ \emph {et~al.}(2015)\citenamefont
  {Maschek}, \citenamefont {Rosenkranz}, \citenamefont {Heid}, \citenamefont
  {Said}, \citenamefont {Giraldo-Gallo}, \citenamefont {Fisher},\ and\
  \citenamefont {Weber}}]{maschek2015wave}%
  \BibitemOpen
  \bibfield  {author} {\bibinfo {author} {\bibfnamefont {M.}~\bibnamefont
  {Maschek}}, \bibinfo {author} {\bibfnamefont {S.}~\bibnamefont {Rosenkranz}},
  \bibinfo {author} {\bibfnamefont {R.}~\bibnamefont {Heid}}, \bibinfo {author}
  {\bibfnamefont {A.~H.}\ \bibnamefont {Said}}, \bibinfo {author}
  {\bibfnamefont {P.}~\bibnamefont {Giraldo-Gallo}}, \bibinfo {author}
  {\bibfnamefont {I.}~\bibnamefont {Fisher}}, \ and\ \bibinfo {author}
  {\bibfnamefont {F.}~\bibnamefont {Weber}},\ }\href@noop {} {\bibfield
  {journal} {\bibinfo  {journal} {Physical Review B}\ }\textbf {\bibinfo
  {volume} {91}},\ \bibinfo {pages} {235146} (\bibinfo {year}
  {2015})}\BibitemShut {NoStop}%
\bibitem [{\citenamefont {Moore}\ \emph {et~al.}(2010)\citenamefont {Moore},
  \citenamefont {Brouet}, \citenamefont {He}, \citenamefont {Lu}, \citenamefont
  {Ru}, \citenamefont {Chu}, \citenamefont {Fisher},\ and\ \citenamefont
  {Shen}}]{moore2010ErTe3}%
  \BibitemOpen
  \bibfield  {author} {\bibinfo {author} {\bibfnamefont {R.~G.}\ \bibnamefont
  {Moore}}, \bibinfo {author} {\bibfnamefont {V.}~\bibnamefont {Brouet}},
  \bibinfo {author} {\bibfnamefont {R.}~\bibnamefont {He}}, \bibinfo {author}
  {\bibfnamefont {D.~H.}\ \bibnamefont {Lu}}, \bibinfo {author} {\bibfnamefont
  {N.}~\bibnamefont {Ru}}, \bibinfo {author} {\bibfnamefont {J.-H.}\
  \bibnamefont {Chu}}, \bibinfo {author} {\bibfnamefont {I.~R.}\ \bibnamefont
  {Fisher}}, \ and\ \bibinfo {author} {\bibfnamefont {Z.-X.}\ \bibnamefont
  {Shen}},\ }\href@noop {} {\bibfield  {journal} {\bibinfo  {journal} {Physical
  Review B}\ }\textbf {\bibinfo {volume} {81}},\ \bibinfo {pages} {073102}
  (\bibinfo {year} {2010})}\BibitemShut {NoStop}%
\bibitem [{\citenamefont {Ru}\ and\ \citenamefont
  {Fisher}(2006)}]{ru2006growth}%
  \BibitemOpen
  \bibfield  {author} {\bibinfo {author} {\bibfnamefont {N.}~\bibnamefont
  {Ru}}\ and\ \bibinfo {author} {\bibfnamefont {I.~R.}\ \bibnamefont
  {Fisher}},\ }\href@noop {} {\bibfield  {journal} {\bibinfo  {journal}
  {Physical Review B}\ }\textbf {\bibinfo {volume} {73}},\ \bibinfo {pages}
  {033101} (\bibinfo {year} {2006})}\BibitemShut {NoStop}%
\bibitem [{\citenamefont {Ruff}\ \emph {et~al.}(2012)\citenamefont {Ruff},
  \citenamefont {Chu}, \citenamefont {Kuo}, \citenamefont {Das}, \citenamefont
  {Nojiri}, \citenamefont {Fisher},\ and\ \citenamefont
  {Islam}}]{ruff2012susceptibility}%
  \BibitemOpen
  \bibfield  {author} {\bibinfo {author} {\bibfnamefont {J.~P.~C.}\
  \bibnamefont {Ruff}}, \bibinfo {author} {\bibfnamefont {J.-H.}\ \bibnamefont
  {Chu}}, \bibinfo {author} {\bibfnamefont {H.-H.}\ \bibnamefont {Kuo}},
  \bibinfo {author} {\bibfnamefont {R.~K.}\ \bibnamefont {Das}}, \bibinfo
  {author} {\bibfnamefont {H.}~\bibnamefont {Nojiri}}, \bibinfo {author}
  {\bibfnamefont {I.~R.}\ \bibnamefont {Fisher}}, \ and\ \bibinfo {author}
  {\bibfnamefont {Z.}~\bibnamefont {Islam}},\ }\href@noop {} {\bibfield
  {journal} {\bibinfo  {journal} {Physical Review Letters}\ }\textbf {\bibinfo
  {volume} {109}},\ \bibinfo {pages} {027004} (\bibinfo {year}
  {2012})}\BibitemShut {NoStop}%
\bibitem [{\citenamefont {Zapf}\ \emph {et~al.}(2014)\citenamefont {Zapf},
  \citenamefont {Stingl}, \citenamefont {Post}, \citenamefont {Maiwald},
  \citenamefont {Bach}, \citenamefont {Pietsch}, \citenamefont {Neubauer},
  \citenamefont {L{\"o}hle}, \citenamefont {Clauss}, \citenamefont {Jiang},
  \citenamefont {Jeevan}, \citenamefont {Basov}, \citenamefont {Gegenwart},\
  and\ \citenamefont {Dressel}}]{zapf2014persistent}%
  \BibitemOpen
  \bibfield  {author} {\bibinfo {author} {\bibfnamefont {S.}~\bibnamefont
  {Zapf}}, \bibinfo {author} {\bibfnamefont {C.}~\bibnamefont {Stingl}},
  \bibinfo {author} {\bibfnamefont {K.~W.}\ \bibnamefont {Post}}, \bibinfo
  {author} {\bibfnamefont {J.}~\bibnamefont {Maiwald}}, \bibinfo {author}
  {\bibfnamefont {N.}~\bibnamefont {Bach}}, \bibinfo {author} {\bibfnamefont
  {I.}~\bibnamefont {Pietsch}}, \bibinfo {author} {\bibfnamefont
  {D.}~\bibnamefont {Neubauer}}, \bibinfo {author} {\bibfnamefont
  {A.}~\bibnamefont {L{\"o}hle}}, \bibinfo {author} {\bibfnamefont
  {C.}~\bibnamefont {Clauss}}, \bibinfo {author} {\bibfnamefont
  {S.}~\bibnamefont {Jiang}}, \bibinfo {author} {\bibfnamefont {H.~S.}\
  \bibnamefont {Jeevan}}, \bibinfo {author} {\bibfnamefont {D.~N.}\
  \bibnamefont {Basov}}, \bibinfo {author} {\bibfnamefont {P.}~\bibnamefont
  {Gegenwart}}, \ and\ \bibinfo {author} {\bibfnamefont {M.}~\bibnamefont
  {Dressel}},\ }\href@noop {} {\bibfield  {journal} {\bibinfo  {journal}
  {Physical Review Letters}\ }\textbf {\bibinfo {volume} {113}},\ \bibinfo
  {pages} {227001} (\bibinfo {year} {2014})}\BibitemShut {NoStop}%
\bibitem [{\citenamefont {Ando}, \citenamefont {Lavrov},\ and\ \citenamefont
  {Komiya}(2003)}]{ando2003anisotropic}%
  \BibitemOpen
  \bibfield  {author} {\bibinfo {author} {\bibfnamefont {Y.}~\bibnamefont
  {Ando}}, \bibinfo {author} {\bibfnamefont {A.~N.}\ \bibnamefont {Lavrov}}, \
  and\ \bibinfo {author} {\bibfnamefont {S.}~\bibnamefont {Komiya}},\
  }\href@noop {} {\bibfield  {journal} {\bibinfo  {journal} {Physical review
  letters}\ }\textbf {\bibinfo {volume} {90}},\ \bibinfo {pages} {247003}
  (\bibinfo {year} {2003})}\BibitemShut {NoStop}%
\bibitem [{\citenamefont {Chu}\ \emph {et~al.}(2010{\natexlab{b}})\citenamefont
  {Chu}, \citenamefont {Analytis}, \citenamefont {Press}, \citenamefont
  {De~Greve}, \citenamefont {Ladd}, \citenamefont {Yamamoto},\ and\
  \citenamefont {Fisher}}]{chu2010plane}%
  \BibitemOpen
  \bibfield  {author} {\bibinfo {author} {\bibfnamefont {J.-H.}\ \bibnamefont
  {Chu}}, \bibinfo {author} {\bibfnamefont {J.~G.}\ \bibnamefont {Analytis}},
  \bibinfo {author} {\bibfnamefont {D.}~\bibnamefont {Press}}, \bibinfo
  {author} {\bibfnamefont {K.}~\bibnamefont {De~Greve}}, \bibinfo {author}
  {\bibfnamefont {T.~D.}\ \bibnamefont {Ladd}}, \bibinfo {author}
  {\bibfnamefont {Y.}~\bibnamefont {Yamamoto}}, \ and\ \bibinfo {author}
  {\bibfnamefont {I.~R.}\ \bibnamefont {Fisher}},\ }\href@noop {} {\bibfield
  {journal} {\bibinfo  {journal} {Physical Review B}\ }\textbf {\bibinfo
  {volume} {81}},\ \bibinfo {pages} {214502} (\bibinfo {year}
  {2010}{\natexlab{b}})}\BibitemShut {NoStop}%
\bibitem [{\citenamefont {Riggs}\ \emph {et~al.}(2015)\citenamefont {Riggs},
  \citenamefont {Shapiro}, \citenamefont {Maharaj}, \citenamefont {Raghu},
  \citenamefont {Bauer}, \citenamefont {Baumbach}, \citenamefont
  {Giraldo-Gallo}, \citenamefont {Wartenbe},\ and\ \citenamefont
  {Fisher}}]{riggs2015URS}%
  \BibitemOpen
  \bibfield  {author} {\bibinfo {author} {\bibfnamefont {S.~C.}\ \bibnamefont
  {Riggs}}, \bibinfo {author} {\bibfnamefont {M.~C.}\ \bibnamefont {Shapiro}},
  \bibinfo {author} {\bibfnamefont {A.~V.}\ \bibnamefont {Maharaj}}, \bibinfo
  {author} {\bibfnamefont {S.}~\bibnamefont {Raghu}}, \bibinfo {author}
  {\bibfnamefont {E.~D.}\ \bibnamefont {Bauer}}, \bibinfo {author}
  {\bibfnamefont {R.~E.}\ \bibnamefont {Baumbach}}, \bibinfo {author}
  {\bibfnamefont {P.}~\bibnamefont {Giraldo-Gallo}}, \bibinfo {author}
  {\bibfnamefont {M.}~\bibnamefont {Wartenbe}}, \ and\ \bibinfo {author}
  {\bibfnamefont {I.~R.}\ \bibnamefont {Fisher}},\ }\href@noop {} {\bibfield
  {journal} {\bibinfo  {journal} {Nature Communications}\ }\textbf {\bibinfo
  {volume} {6}} (\bibinfo {year} {2015})}\BibitemShut {NoStop}%
\end{thebibliography}%

\end{document}